\documentclass[fdp,a4paper,fleqn%
]{w-art}
\usepackage{times,cite,w-thm,amssymb}
\theoremstyle{plain}

\theoremstyle{definition}

\usepackage[]{graphicx}

\usepackage{color} 
\usepackage{multirow} 
\usepackage[T1]{fontenc} 
\usepackage{slashed}

\newcommand{\be}{\begin{equation}}
\newcommand{\ee}{\end{equation}}
\newcommand{\bea}{\begin{eqnarray}}
\newcommand{\eea}{\end{eqnarray}}

\let\vev\VEV

\def\e6{$E(6)$}
\def\10{$SO(10)$}
\def\21{$SU(2) \otimes U(1) $}

\def\422{$SU(4) \otimes SU(2) \otimes SU(2)$ }
\def\321{$SU(3) \otimes SU(2) \otimes U(1)$ }
\def\lfv{lepton flavour violation }
\def\lnv{lepton number violation }
\def\O{\hbox{$\cal O$ }}

\def \znbb {$0\nu\beta\beta$ }

\def\lsim{\raise0.3ex\hbox{$\;<$\kern-0.75em\raise-1.1ex\hbox{$\sim\;$}}}
\def\gsim{\raise0.3ex\hbox{$\;>$\kern-0.75em\raise-1.1ex\hbox{$\sim\;$}}}
\def\vev#1{\left\langle #1\right\rangle}

\begin{document}
\DOIsuffix{theDOIsuffix}
\Volume{55}
\Month{01}
\Year{2007}
\pagespan{1}{}
\Receiveddate{XXXX}
\Reviseddate{XXXX}
\Accepteddate{XXXX}
\Dateposted{XXXX}



\title[Neutrino masses and mixing: a
   flavour symmetry roadmap]{Neutrino masses and mixing: a
   flavour symmetry roadmap}


 \author[S.\,Morisi]{S.\,Morisi\inst{1,2,}\footnote{Corresponding
     author\quad E-mail:~\textsf{morisi@ific.uv.es}, Phone:
     +34 963543519, Fax: +34 963543488}}
 \author[J.~W.~F.~Valle]{J.~W.~F.~Valle\inst{1,}%
   \footnote{http://astroparticles.ific.uv.es/}}  \address[\inst{1}]{ \it AHEP
   Group, Instituto de F\'{\i}sica Corpuscular --
   C.S.I.C./Universitat de Val{\`e}ncia \\
   Edificio de Institutos de Paterna, Apartado 22085, E--46071
   Val{\`e}ncia, Spain}

\begin{abstract}

  Over the last ten years tri-bimaximal mixing has played an important
  role in modeling the flavour problem.  We give a short review of the
  status of flavour symmetry models of neutrino mixing.  We
  concentrate on non-Abelian discrete symmetries, which provide a
  simple way to account for the TBM pattern. 
  We discuss phenomenological
  implications such as neutrinoless double beta decay, lepton flavour
  violation as well as theoretical aspects such as the possibility to
  explain quarks and leptons within a common framework, such as grand
  unified models.

\end{abstract}
\maketitle                   





 \tableofcontents  

\section{Introduction}

The long-standing solar and atmospheric neutrino anomalies suggested
the idea of neutrino oscillations, now confirmed by a series of
``laboratory'' experiments based on reactors and
accelerators~\cite{art:2012}.
Especially puzzling theoretically is the fact that the neutrino mixing
angles inferred from experiment follow a pattern rather different from
that which characterizes quark
mixing~\cite{maltoni:2004ei,schwetz:2008er}. The atmospheric angle
$\theta_{23}$ is close to maximal, with a large value for the solar
angle $\theta_{12}$, both of which are at odds with their quark
sector counterparts.
Moreover, following the first indications of nonzero $\theta_{13}$
reported by accelerator experiments MINOS~\cite{Adamson:2011qu} and
T2K~\cite{Abe:2011sj} three recent measurements of $\theta_{13}$ have
been reported by the reactor experiments Double
CHOOZ~\cite{Abe:2011fz}, Daya Bay~\cite{An:2012eh} and
RENO~\cite{Ahn:2012nd}, as well as by the MINOS
collaboration~\cite{nichol-nu2012}~\footnote{ The bulk of the data on
  neutrino oscillations are well described in terms of three active
  neutrinos.}.

While the historic discovery of neutrino oscillations provides strong
indications for the need of physics beyond Standard Model (SM), the
detailed nature of this physics remains elusive: i) the mechanism
responsible for neutrino mass generation, ii) its flavour structure,
iii) its characteristic scale, as well as the nature of the associated
messenger particle all remain unknown.
As a result the nature of neutrinos, their mass and mixing parameters
are so far unpredicted~\cite{nunokawa:2007qh}.

Understanding the pattern of neutrino mixing is part of the flavour
problem, one of the deepest in particle physics. Although it may be the
result of an accident, the pattern of neutrino mixing angles most
likely follows a \textit{rationale}.
Indeed there has been a strong effort towards the formulation of
symmetry--based approaches to address the flavour problem from first
principles, assuming the existence of an underlying flavour symmetry of
leptons and/or quarks, separately or jointly.

\vskip3.mm

In 2002 Harrison, Perkins and Scott proposed the tri-bimaximal (TBM)
mixing \emph{ansatz}~\cite{Harrison:2002er} with effective bimaximal
mixing of $\nu_{\mu}$ and $\nu_{\tau}$ at the atmospheric scale and
effective trimaximal mixing for $\nu_e$ with $\nu_{\mu}$ and
$\nu_{\tau}$ at the solar scale (hence `tri-bimaximal' mixing).
While large atmospheric mixing was already discussed before 2002, the
trimaximal solar angle has represented a milestone for model
building. Non-Abelian continuous and discrete flavour symmetries have
been extensively used to account for TBM mixing. Here we review the
basic features of some of the most interesting models proposed in the
last ten years.

To be fair we must say that the global analysis of neutrino
oscillation data now indicates a robust measurement of a relatively
``large'' value of $\theta_{13}$~\cite{Tortola:2012te} which casts
some doubt on the validity of the TBM \emph{ansatz} as a good first
approximation to the neutrino mixing pattern. Nevertheless it is too
early to jump into conclusions, since in concrete theories there may
be large corrections to the TBM pattern, so that here we still take it
as a useful reference \emph{ansatz}.

Non-Abelian discrete groups have non trivial irreducible
representations ({\it irreps}). Assigning the three known generations
of leptons to irreps of a flavour symmetry group one can make
predictions for masses and mixings in the lepton sector. 
In general it is expected that the number of free parameters of models
based on Abelian flavour symmetries is typically larger then the
corresponding number of free parameters needed to describe non-Abelian
flavour symmetry models. Moreover there are non-Abelian discrete groups
that contain triplet irreps, exactly as the number of generations in
the standard model. Hence there are viable and predictive non-Abelian
models to which we dedicate this brief review.
 
The smallest group~\footnote{The order of a finite group is just the
  number of elements.} that contains triplet irreps is $A_4$, the
group of the even permutations of four objects, isomorphic to the
group of the symmetries of the tetrahedron $T$.  For a classification
of the irreps of different non Abelian discrete groups see for
instance \cite{Ishimori:2010au}.  $A_4$ was first used in the lepton
sector by Ma and Rajasekaran \cite{Ma:2001dn} but the solar angle was
not predicted and neutrino masses were degenerate.  A realistic model
was proposed by Babu, Ma and Valle \cite{Babu:2002dz} adopting a
supersymmetric context in order to produce the required neutrino mass
splittings and mixing angles, predicting maximal atmospheric mixing
and vanishing $\theta_{13}$ to first approximation. Although expected
to be sizeable, the solar mixing angle is unpredicted. 
In order to predict the full tri-bimaximal pattern, the neutrino mass
matrix must take the form 
\begin{equation}\label{mTBM}
M_\nu=
\left(
\begin{array}{ccc}
y&x&x\\
x&y+z&x-z\\
x&y-z&x+z\\
\end{array}
\right),
\end{equation} 
where $x,y,z$ are free parameters. The above matrix has two properties: 
\begin{itemize}
\item it is $\mu-\tau$ invariant giving maximal atmospheric and zero
  reactor angles;
\item it satisfies the relation
  $(M_{\nu})_{11}+(M_{\nu})_{12}=(M_{\nu})_{22}+(M_{\nu})_{23}$ giving
  trimaximal solar agle.
\end{itemize}
The neutrino mass matrix of eq.\,(\ref{mTBM}) is diagonalized by the
TBM mixing matrix 
\begin{equation}\label{UTBM}
U =
\left(
\begin{array}{ccc}
2/\sqrt{6}  & 1/\sqrt{3}& 0\\
-1/\sqrt{6}  & 1/\sqrt{3} & 1/\sqrt{2} \\
-1/\sqrt{6}  & 1/\sqrt{3}& -1/\sqrt{2}\\
\end{array}
\right),
\end{equation} 
independently of the mass eigenvalues.  A trimaximal solar angle was
first given in a paper of Ma~\cite{Ma:2004zv} based on type-II seesaw
but assuming in an {\it ad hoc } way that the contribution of two
scalar singlets ${\bf 1}$ and ${\bf 1'}$ were the same.  The
derivation of TBM mixing from a flavour symmetry was achieved by
Altarelli and Feruglio in Refs.~\cite{Altarelli:2005yp} and
\cite{Altarelli:2005yx} and subsequently by Babu and He
\cite{Babu:2005se}.

In contrast to the quark sector, neutrino mixing angles are large,
possibly TBM, since the flavour group can break into two different
subgroups in the charged and neutral lepton sectors, respectively.
Consider $A_4$ as an example. $A_4$ contains two abelian subgroups,
namely $Z_2$ and $Z_3$.  When broken into $Z_3$ in the charged sector,
and into $Z_2$ in the neutrino sector, $A_4$ leads to a lepton mixing
matrix of TBM form, Eq.~(\ref{UTBM}).

Of course $A_4$ is totally broken, therefore deviations at next to
leading order are expected.  In general one can not align $A_4$ in the
$Z_3$ and $Z_2$ directions in the charged and neutral lepton sectors
respectively, this is known as the {\it alignment} problem.
This may be circumvented by using extra dimensions and/or
supersymmetry \cite{Altarelli:2005yp,Altarelli:2005yx} or by assuming
a suitably chosen soft breaking sector.  Alternatively, using a large
discrete group, namely $ Z_2^3\times U_L(1)^3 \rtimes S_3$, Grimus and
Lavoura have shown \cite{Grimus:2008vg} how to obtain the TBM form
without alignment problem.
  
\section{The origin of neutrino mass}
\label{sec:origin-neutrino-mass}

Table \ref{tab:sm} lists the fifteen fundamental ``left-handed''
chiral fermions of the Standard Model (SM), sequentially repeated, one
set for each generation.  In contrast to charged fermions, neutrinos
come only in one chiral species, and parity violation in the weak
interaction is introduced explicitly by having only ``left'' fermions
transforming as doublets under the \321 gauge group.

The simplest and most general way to generate neutrino mass in the
Standard \321 Model (SM) is by adding an effective dimension-five
operator \O$_{ab} = \lambda_{ab} \ell_a \ell_b \Phi \Phi$, where
$\ell_a$ denotes any of the three lepton doublets and $\Phi$ is the SM
scalar doublet.  ~\cite{Weinberg:1980bf}.
\begin{figure}[h] \centering
   \includegraphics[angle=90,width=.35\linewidth]{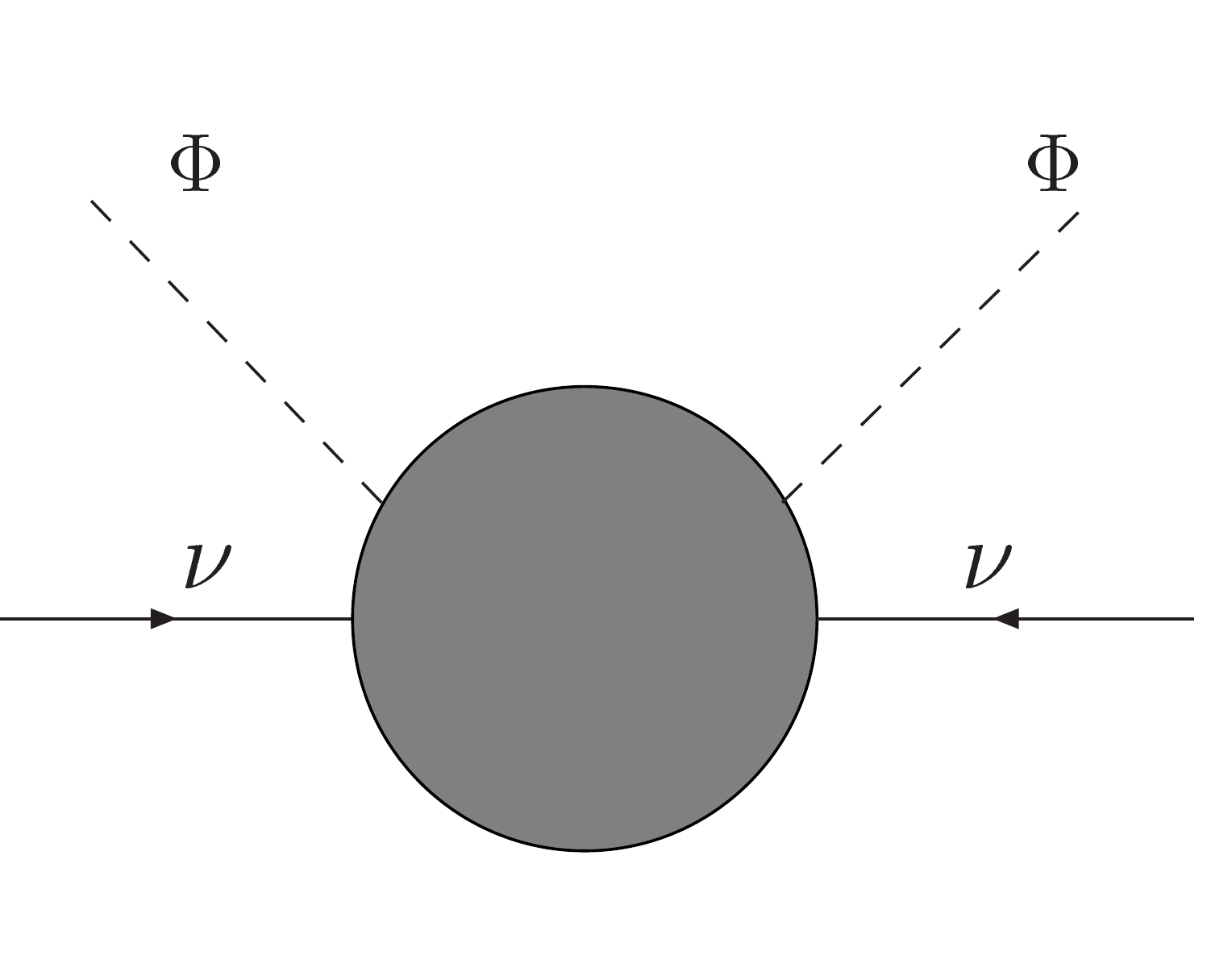}
    \caption{\label{fig:d5} 
    Dimension five  operator yielding neutrino mass.}
\end{figure}
\begin{table*}
  \centering
\begin{math}
\begin{array}{|c|c|} \hline
& \ \ \ {\mbox SU(3)\otimes SU(2)\otimes U(1)} \\
 \hline
\hline
\ell_a = (\nu_a, l_a)^T & (1,2,-1)\\
e_a^c   & (1,1,2)\\
\hline \hline
Q_a = (u_a, d_a)^T    & (3,2,1/3)\\
u_a^c   & (\bar{3},1,-4/3)\\
d_a^c   & (\bar{3},1,2/3)\\
\hline \hline
\Phi  & (1,2,1)\\
\hline
\end{array}
\end{math}
\caption{Lepton, quark and scalar multiplets of the Standard Model}
\label{tab:sm}
\end{table*}
After electroweak symmetry breaking takes place, through the nonzero
vacuum expectation value (vev) $\vev{\Phi}$, Majorana neutrino masses
are induced.  From such general point of view the emergence of Dirac
neutrinos would be an ``accident'', justified only in the presence of
a fundamental lepton number symmetry, in general absent.
The underlying nature of the dimension five operator in
Fig.~\ref{fig:d5} is unknown: little can be said from first
principles about the {\sl mechanism} that engenders \O$_{ab}$, its
associated mass {\sl scale} or its {\sl flavour structure}.
The strength of the operator \O$_{ab}$ can be naturally suppressed if
the associated messengers are superheavy, as expected say, in unified
scenarios. Alternatively, its strength can be naturally suppressed
even in the absence of heavy messengers, due to the fact that
\O$_{ab}$ violates lepton number by two units ($\Delta L=2$), i.e. in
its absence the theory recovers lepton number conservation.
This is known as t'Hofft's naturalness~\cite{tHooft:1979bh}.
Correspondingly, one may have high and low-scale neutrino mass models,
depending on the mass characterizing the messengers whose exchange
induces \O$_{ab}$.
While the former type are closer to the idea of unification, the
latter are closer to experimental testability.

\subsection{High scale seesaw mechanisms}
\label{sec:high-scale-seesaw}

\begin{figure}[ht] \centering
 \includegraphics[height=3.5cm,width=.7\linewidth]{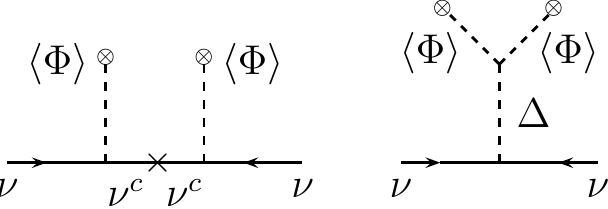}
    \caption{\label{fig:seesaw} %
      Left: type-I seesaw  (right-handed neutrino
      exchange). Right:  type-II seesaw (scalar triplet 
      exchange).}
\end{figure}

The exchange of heavy messenger states, either fermions (type-I or
type-III seesaw) or scalars (type-II seesaw) provides a simple way
to generate the operator \O$_{ab}$.
The smallness of its strength is ascribed to the large mass scale
characterizing the violation of total lepton
number~\cite{valle:2006vb,th-talk2}.
The simplest and most general description of the seesaw mechanism is
in terms of just the \321 gauge group with ungauged lepton number
broken either explicitly~\cite{schechter:1980gr} or
spontaneously~\cite{schechter:1982cv}. The latter framework or
``1-2-3'' scheme is characterized by \321 singlet, doublet and triplet
mass terms, described by the
matrix~\cite{schechter:1980gr,schechter:1982cv}
\be
\label{ss-matrix} {\mathcal M_\nu} = \left(\begin{array}{cc}
    Y_3 v_3 & Y_\nu  v_2 \\
    {Y_\nu}^{T} v_2  & Y_1 v_1 \\
\end{array}\right) 
\ee 
where $v_2 \equiv \vev{\Phi}$ denotes the SM Higgs doublet vev and the
basis is $\nu_{L}$, $\nu^{c}_{L}$, corresponding to the three ``left''
and three ``right'' neutrinos, respectively. Note that, though
symmetric, by the Pauli principle, ${\mathcal M_\nu}$ is complex, so
that its Yukawa coupling sub-matrices \(Y_\nu\) as well as \(Y_3\) and
\(Y_1\) are complex matrices denoting the relevant Yukawa couplings,
the last two symmetric.
Such \321 seesaw contains singlet, doublet and triplet scalar
multiplets, obeying a simple ``1-2-3'' vev--seesaw relation of the type
 \begin{equation}
   v_3 v_1 \sim {v_2}^2 \:\:\: \mathrm{with} \:\:\: v_1 \gg v_2 \gg v_3 
 \label{eq:123-vev-seesaw}
 \end{equation}
 This vev--seesaw is consistent with the minimization condition of the
 \321 invariant scalar potential, and implies that the triplet vev
 $v_3 \to 0$ as the singlet vev $v_1$ grows.
 Neutrino masses are suppressed either by heavy \321 singlet
 ``right-handed'' neutrino exchange (type I) or by the smallness of
 the induced triplet vev that follows from heavy scalar exchange (type
 II), as illustrated in Fig.~\ref{fig:seesaw}.
The matrix \(\mathcal{M_\nu}\) is diagonalized by a unitary mixing
matrix \(U_\nu\),
\begin{equation}
   U_\nu^T {\mathcal M_\nu} U_\nu = \mathrm{diag}(m_i,M_i),
\end{equation}
yielding 6 mass eigenstates: the three light neutrinos with
masses \(m_i\), and the three heavy two-component leptons.
The light neutrino mass states \(\nu_i\) are given in terms of the
flavour eigenstates via the unitary matrix
$U_\nu$~\cite{schechter:1980gr}
\begin{equation}
   \nu_i = \sum_{a=1}^{6}(U_\nu)_{ia} n_a.
\end{equation}
where the diagonalization matrices are given as a perturbation series,
see Ref.~\cite{schechter:1982cv}.
The effective light neutrino mass, obtained this way is of the
form
\begin{equation}
  \label{eq:ss-formula0}
  m_{\nu} \simeq Y_3 v_3 -
Y_\nu {Y_1}^{-1} {Y_\nu}^T \frac{{v_2}^2}{v_1}~.
\end{equation}

Since in such ``1-2-3'' seesaw lepton number is ungauged, there is a
physical Goldstone boson resulting from its spontaneous breakdown,
namely the Majoron~\cite{chikashige:1981ui,schechter:1982cv}.
It is often argued that, due to quantum gravity effects the associated
Majoron will pick up a mass. It has been shown that, a keV range
Majoron can provide the observed dark matter of the
Universe~\cite{Lattanzi:2007ux} and be detected through its X-ray
gamma line searches~\cite{bazzocchi:2008fh}.
If B-L is gauged~\cite{valle:1987sq} the Majoron is absorbed as the
longitudinal mode of a new neutral gauge boson.

\subsection{Low-scale seesaw mechanisms}
\label{sec:low-scale-seesaw}

A distinguishing feature of the seesaw mechanism as proposed in
Ref.~\cite{schechter:1980gr,schechter:1982cv,valle:2006vb} and other
presentations~\cite{th-talk2} is that it is formulated in terms of the
standard \321 SM gauge group. The higher generality implies that the
number of ``right-handed'' neutrinos is totally arbitrary since, being
gauge singlets, they carry no anomaly.  New important features may
emerge when the seesaw is realized with non-minimal lepton content,
opening the door to the possibility of low-scale seesaw mechanisms,
such as the inverse seesaw~\cite{mohapatra:1986bd}.

\subsubsection{Inverse seesaw mechanisms}
\label{sec:inverse-sees-mech}

\begin{figure}[h] \centering
    \includegraphics[height=3cm,width=.45\linewidth]{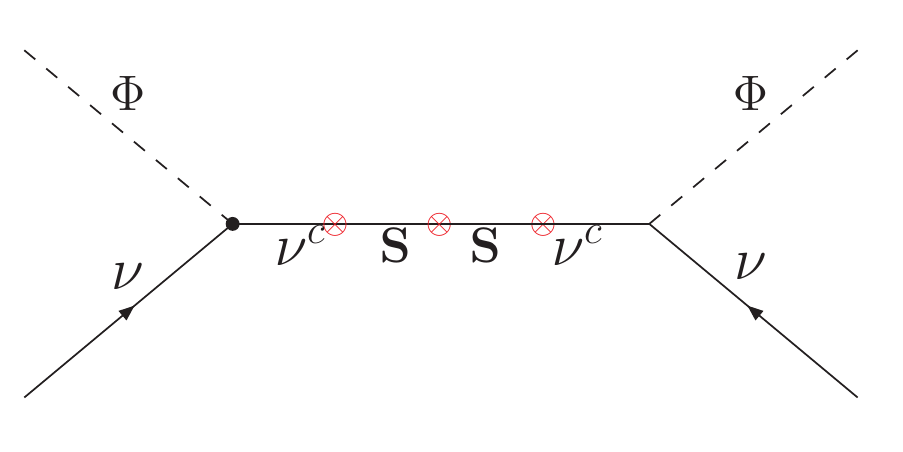}
    \caption{\label{fig:iss-mass} 
    Inverse seesaw mechanism.}
\end{figure}

The model adds a pair of two-component \321 singlet leptons, $\nu_i^c,
S_i$, to each SM generation \(i\) running over \(1,2,3\).
In the basis \(\nu,\nu^c,S\), the neutral leptons mass matrix
$\mathcal{M_\nu}$ is \(9\times9\), i.e.
\begin{equation}
\label{eqn:doubleSeesaw}
{\mathcal M_\nu}=\left(
   \begin{array}{ccc}
      0   & Y_\nu^T v_2 & 0   \\
      Y_\nu  v_2 & 0     & M^T \\
      0   & M     & \mu
   \end{array}\right),
\end{equation}
with \(\mu \ll Y_\nu v_2 \ll M\), where \(Y_\nu\) and \(M\) are
arbitrary \(3\times3\) complex Yukawa matrices, while \(\mu\) is
complex symmetric, due to the Pauli principle.
In such a scheme the three light neutrino masses are determined from
the effective  \(3\times3\) neutrino mass matrix
\begin{equation}
\label{eqn:lightNu}
    m_\nu \approx  {{v_2}^2} Y_\nu^T { M^{T}}^{-1} \mu M^{-1}  Y_\nu~.
\end{equation}
The mass generation is illustrated in Fig.~\ref{fig:iss-mass}.
Notice that as \(\mu\) $\to$ 0 all neutrinos become massless and
lepton number symmetry is restored.
The entry \(\mu\) may be proportional to the vev of an \321 singlet
scalar, in which case the model contains a singlet
Majoron~\cite{gonzalez-garcia:1989rw} which may provide an invisible
Higgs boson decay channel~\cite{joshipura:1992hp}.  In such schemes
one must take into account the existence of sizeable invisible Higgs
boson decay channels in the analysis of experimental data on Higgs
searches~\cite{deCampos:1997bg,Abdallah:2003ry}.


\subsubsection{Linear seesaw mechanism}
\label{sec:linear-seesaw}

\begin{figure}[h] \centering
    \includegraphics[height=3cm,width=.45\linewidth]{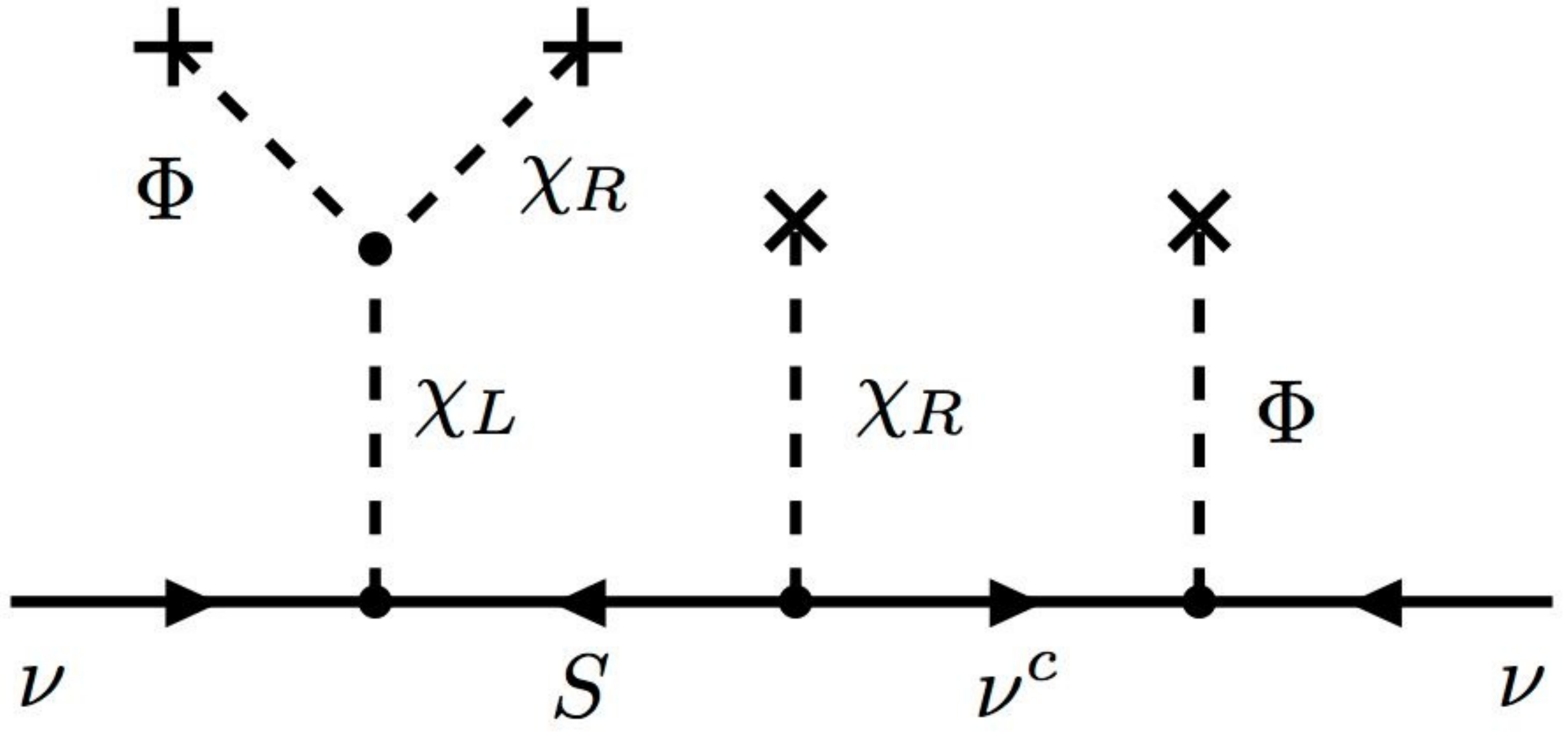}
    \caption{\label{fig:new-seesaw} 
    Linear seesaw mechanism.}
\end{figure}

We now turn to a low-energy seesaw mechanism with gauged B-L,
originally suggested in the framework of dynamical left-right
symmetry~\cite{akhmedov:1995ip,akhmedov:1995vm} and more recently in
an \10 model with broken D-parity in which only an Abelian factor
survives at low energies~\cite{malinsky:2005bi}.
In addition to the three left- and right-handed neutrinos the model
contains three sequential gauge singlets $S_{iL}$ with the following
mass matrix
\be \label{ess-matrix}
{\mathcal M_\nu} =
\left(\begin{array}{ccc}
0 & Y_\nu \vev{\Phi} & F \vev{\chi_L} \\
{Y_\nu}^{T} \vev{\Phi} & 0 & \tilde F \vev{\chi_R}     \\
F^{T} \vev{{\chi}_L}    & \tilde F^{T} \vev{\chi_R} & 0
\end{array}\right) 
\ee 
in the basis $\nu_{L}$, $\nu^{c}_{L}$, $S_{L}$. 
The zeros along the diagonal, in the $\nu_{L}$-$\nu_{L}$ and
$\nu^{c}_{L}$-$\nu^{c}_{L}$ entries, are due to the fact that there is
no {\bf 126}. The resulting neutrino mass is 
\begin{eqnarray}
m_{\nu} & \simeq &
\frac{\vev{\Phi}^2}{M_\mathrm{unif}} 
\left[Y_\nu ( F \tilde F^{-1})^{T}+( F \tilde F^{-1}) {Y_\nu}^{T}\right],
\end{eqnarray}
where $M_\mathrm{unif}$ is the unification scale, $F$ and $\tilde F$
denote independent Yukawa coupling combinations of the $S_{iL}$.
One can see that the neutrino mass is suppressed by the unification
scale $M_{\mathrm{unif}}$ \emph{instead of the B-L breaking
  scale}.  
Note that, in contrast to other seesaw schemes, this one is {\sl
  linear} in the Dirac Yukawa couplings $Y_\nu$, as illustrated in
Fig.~\ref{fig:new-seesaw}.
It is rather remarkable that one can indeed take the B-L scale as low
as TeV without generating inconsistencies with gauge coupling
unification~\cite{malinsky:2005bi}. The light neutral gauge boson may
be searched directly at the Large Hadron Collider (LHC) or through
precision studies of low-energy neutrino-electron
scattering~\cite{Garces:2011aa}.

\subsection{Radiative models of neutrino mass}
\label{sec:radiative-models}

\begin{figure}[h] \centering
   \includegraphics[height=3cm,width=.45\linewidth]{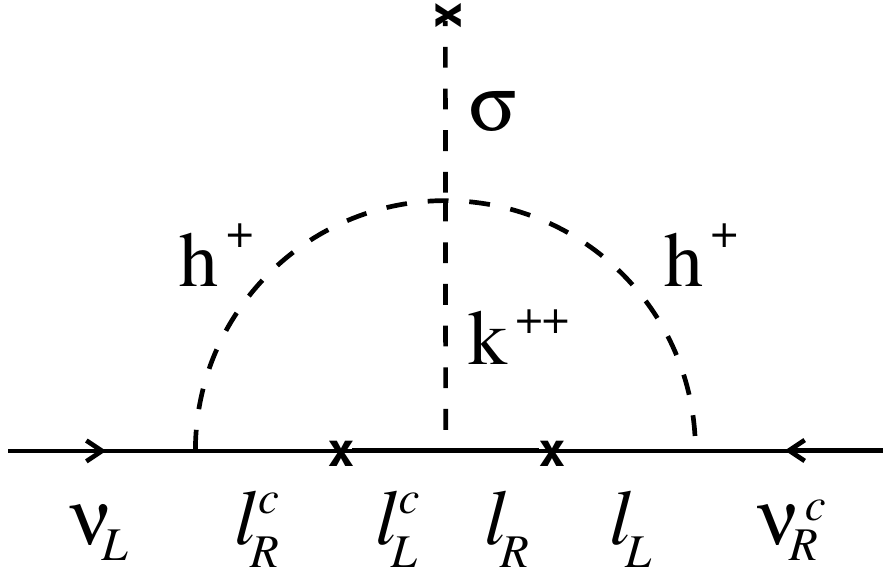}
    \caption{\label{fig:neumass} 
    Two-loop origin for neutrino mass.}
\end{figure}

In addition to the above low-scale seesaw schemes there is a variety
of other models of neutrino mass where the operator \O$_{ab}$ is
induced from physics at accessible scales, TeV or less. 
The first possibility is that neutrino masses are induced by
calculable radiative corrections~\cite{zee:1980ai,babu:1988ki}, for
instance, as illustrated in Fig.~\ref{fig:neumass}. Up to a
logarithmic factor one has, schematically,
\begin{equation}
   \label{eq:babu}
{\mathcal M_\nu} \sim \lambda_0 \left(\frac{1}{16\pi^2}\right)^2 
f Y_l h Y_l f^T \frac{v_2^2}{(m_k)^2} \vev{\sigma}
 \end{equation}
 in the limit where the doubly-charged scalar $k$ is much heavier than
 the singly charged one. Here $l$ denotes a charged lepton, $f$ and
 $h$ are their Yukawa coupling matrices and $Y_l$ denotes the SM Higgs
 Yukawa couplings to charged leptons and $\vev{\sigma}$ is an \321
 singlet vev introduced in Ref.~\cite{Peltoniemi:1993pd}. 
 The smallness of the neutrino mass arises from the presence of a
 product of five small Yukawas and the appearance of the two-loop
 factor. 
 A special feature of the model is that, thanks to the anti-symmetry
 of the $f$ Yukawa coupling matrix, one of the neutrinos is massless.

\subsection{Supersymmetric neutrino masses}
\label{sec:supersymm-as-orig}

An interesting alternative are models where low energy supersymmetry
is the origin of neutrino mass~\cite{Hirsch:2004he} through the
breaking of the so-called R parity. This could arise spontaneously,
driven by a nonzero vev of an \321 singlet
sneutrino~\cite{Masiero:1990uj,romao:1992vu,romao:1997xf}.  This way
we are led to the minimal way to include neutrino masses into the
MSSM, which we take as reference model, with effective bilinear R
parity violation~\cite{Diaz:1998xc}.  
The neutrino mass generation scenario is hybrid, with one scale
generated at tree level by the mixing of neutralinos and neutrinos,
and the other induced by ``calculable'' loop
corrections~\cite{Hirsch:2000ef,diaz:2003as}. 
\begin{figure}[h] \centering
    \includegraphics[height=2.5cm,width=.5\linewidth]{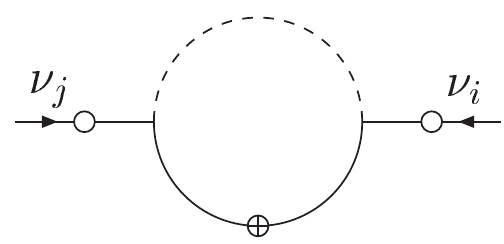}
  \caption{\label{fig:d-5} Loop origin of solar mass scale.
    Atmospheric scale arises from tree-level neutralino exchange.}
\end{figure}
The neutrino mass spectrum naturally follows a normal hierarchy, with
the atmospheric scale generated at the tree level with the solar mass
scale arising from calculable loops, as indicated in
Fig.~\ref{fig:d-5}.

\section{Prototype flavour model with tetrahedral symmetry}
\label{sec:bmv-model}

We now turn to models incorporating flavour symmetries, starting with
the BMV (Babu-Ma-Valle) model~\cite{Babu:2002dz}.  The usual quark,
lepton, and Higgs superfields transform under $A_4$ as follows:

\begin{center}
\begin{tabular}{|c|c|c|c|c|c|c|}
\hline
&$\hat Q$ & $\hat L$ & $\hat u^c_1, ~\hat d^c_1, ~\hat e^c_1$ &$ \hat u^c_2, ~\hat d^c_2, ~\hat e^c_2 $&$ \hat u^c_3, ~\hat d^c_3, ~\hat e^c_3$ & $\hat \phi_{1,2}$\\
\hline
$A_4$ & $3$ & $3$ & $1$ &$1'$ & $1''$& $1$\\
\hline
$Z_3$ & $1$ & $1$ & $\omega$ &$\omega$ & $\omega$& $1$\\
\hline
\end{tabular}
\end{center}
The following heavy $SU(2)$ singlet quark, lepton, and Higgs
superfields are also added
\begin{center}
\begin{tabular}{|c|c|c|c|c|c|c|c|c|}
\hline
&$\hat U$&$\hat U^c$ &$\hat D$ &$\hat D^c$&$\hat E$&$\hat E^c$ &$\hat N^c$ &$\hat \chi$\\
\hline
$A_4$ & $3$ & $3$ & $3$ &$3$ & $3$& $3$& $3$& $3$\\
\hline
$Z_3$ & $1$ & $1$ & $1$& $1$&$1$& $1$& $1$&$\omega^2$\\
\hline
\end{tabular}
\end{center}
with $\omega^3 = 1$ and $1 + \omega + \omega^2 = 0$.  The
superpotential of this model is then given by
\begin{eqnarray}
\hat W &=& M_U \hat U_i \hat U^c_i + f_u \hat Q_i \hat U^c_i \hat \phi_2 +
h^u_{ijk} \hat U_i \hat u^c_j \hat \chi_k +
M_D \hat D_i \hat D^c_i + f_d \hat Q_i \hat D^c_i \hat \phi_1 +
h^d_{ijk} \hat D_i \hat d^c_j \hat \chi_k \nonumber \\
&+& M_E \hat E_i \hat E^c_i + f_e \hat L_i \hat E^c_i \hat \phi_1 +
h^e_{ijk} \hat E_i \hat e^c_j \hat \chi_k +
{1 \over 2} M_N \hat N^c_i \hat N^c_i + f_N \hat L_i \hat N^c_i
\hat \phi_2 + \mu \hat \phi_1 \hat \phi_2 \nonumber \\
&+& {1 \over 2} M_\chi \hat \chi_i \hat \chi_i + h_\chi \hat \chi_1
\hat \chi_2 \hat \chi_3,
\end{eqnarray}
with the usual assignment of $R$ parity to distinguish between the
Higgs superfields, i.e. $\hat \phi_{1,2}$ and $\hat \chi_i$, from the
quark and lepton superfields.  The terms $\hat \chi_i \hat N^c_j \hat
N^c_k$, etc. are forbidden by the $Z_3$.  However, $Z_3$ can break
explicitly but softly, by $M_\chi \neq 0$.
The scalar potential involving $\chi_i$ is given by
\begin{equation}
V = |M_\chi \chi_1 + h_\chi \chi_2 \chi_3|^2 + |M_\chi \chi_2 + h_\chi \chi_3
\chi_1|^2 + |M_\chi \chi_3 + h_\chi \chi_1 \chi_2|^2,
\end{equation}
which has the supersymmetric solution $(V=0)$
\begin{equation}
\langle \chi_1 \rangle = \langle \chi_2 \rangle = \langle \chi_3 \rangle =
u = -M_\chi/h_\chi,
\end{equation}
so that the breaking of $A_4$ at the high scale $M_\chi$ does not
break the supersymmetry.  
Consider now the $6 \times 6$ Dirac mass matrix linking $(e_i,E_i)$ to
$(e_j^c,E_j^c)$.
\begin{equation}
{\cal M}_{eE} = \left[ \begin{array} {c@{\quad}c@{\quad}c@{\quad}c@{\quad}c@
{\quad}c} 0 & 0 & 0 & f_e v_1 & 0 & 0 \\ 0 & 0 & 0 & 0 & f_e v_1 & 0 \\
0 & 0 & 0 & 0 & 0 & f_e v_1 \\ h_1^e u & h_2^e u & h_3^e u & M_E & 0 & 0 \\
h_1^e u & h_2^e \omega u & h_3^e \omega^2 u & 0 & M_E & 0 \\
h_1^e u & h_2^e \omega^2 u & h_3^e \omega u & 0 & 0 & M_E \end{array} \right],
\end{equation}
where $v_1 = \langle \phi_1^0 \rangle$ with similar forms for the
quark mass matrices. The reduced $3 \times 3$ charged leptons 
mass matrix is then 
\begin{equation}
{\cal M}_e = U_L \left[ \begin{array} {c@{\quad}c@{\quad}c} {h_1^e}' & 0 & 0
\\ 0 & {h_2^e}' & 0 \\ 0 & 0 & {h_3^e}' \end{array} \right] {\sqrt 3 f_e v_1
u \over M_E},
\end{equation}
where ${h_i^e}' = h_i^e [1+(h_i^e u)^2/M_E^2]^{-1/2}$ and
\begin{equation}
U_L = {1 \over \sqrt 3} \left[ \begin{array} {c@{\quad}c@{\quad}c} 1 & 1 & 1
\\ 1 & \omega & \omega^2 \\ 1 & \omega^2 & \omega \end{array} \right].
\end{equation}
This shows how charged-lepton masses are allowed to be all different,
despite the presence of the $A_4$ symmetry, because there are three
inequivalent one-dimensional representations.  Clearly, the $up$ and
$down$ quark mass matrices are obtained in the same way, both are
diagonalized by $U_L$, so that the charged-current mixing matrix
$V_{CKM}$ is the identity matrix. CKM angles may be generated from
corrections associated to the structure of the soft supersymmetry
breaking sector \cite{hall:1985dx,borzumati:1986qx}.
The $6 \times 6$ Majorana neutrino mass matrix is given by
\begin{equation}
{\cal M}_{\nu N} = \left[ \begin{array} {c@{\quad}c} 0 & U_L f_N v_2
\\ U_L^T f_N v_2 & M_N \end{array} \right],
\end{equation}
in the basis $(\nu_e, \nu_\mu, \nu_\tau, N_1^c, N_2^c, N_3^c)$ and
$v_2 \equiv \langle \phi_2^0 \rangle$.  The effective $(\nu_e,
\nu_\mu, \nu_\tau)$ mass matrix becomes
\begin{equation}
{\cal M}_\nu = {f_N^2 v_2^2 \over M_N} U_L^T U_L = {f_N^2 v_2^2 \over M_N}
\left[ \begin{array} {c@{\quad}c@{\quad}c} 1 & 0 & 0 \\ 0 & 0 & 1 \\
0 & 1 & 0 \end{array} \right].
\end{equation}
showing that neutrino masses are degenerate at this stage.  Consider
now the above as coming from an effective dimension-five operator
\begin{equation}
{f_N^2 \over M_N} \lambda_{ij} \nu_i \nu_j \phi_2^0 \phi_2^0,
\end{equation}
where $\lambda_{ee} = \lambda_{\mu \tau} = \lambda_{\tau \mu} = 1$ and
all other $\lambda$'s are zero, at some high scale.  As we come down
to the electroweak scale, Eq.~(14) is corrected by the wave-function
renormalizations of $\nu_e$, $\nu_\mu$, and $\nu_\tau$, as well as the
corresponding vertex renormalizations, lifting the neutrino degeneracy
due to the different charged-lepton masses.  In order to obtain a
pattern suitable for explaining current neutrino oscillation data we
assume the presence of radiative corrections associated to a general
slepton mass matrix in softly broken
supersymmetry~\cite{Chankowski:2000fp}.  Given the structure of
$\lambda_{ij}$ at the high scale, its low scale form is fixed to first
order as
\begin{equation}
\lambda_{ij} = \left[ \begin{array} {c@{\quad}c@{\quad}c} 1 + 2 \delta_{ee} &
\delta_{e \mu} + \delta_{e \tau} & \delta_{e \mu} + \delta_{e \tau} \\
\delta_{e \mu} + \delta_{e \tau} & 2 \delta_{\mu \tau} & 1 + \delta_{\mu \mu}
+ \delta_{\tau \tau} \\ \delta_{e \mu} + \delta_{e \tau} & 1 + \delta_{\mu \mu}
+ \delta_{\tau \tau} & 2 \delta_{\mu \tau} \end{array} \right],
\end{equation}
where we have assumed all parameters to be real as a first approximation.  
[The above matrix is obtained by multiplying that of Eq.~(13) on the left 
and on the right by all possible $\nu_i \to \nu_j$ transitions.]  Let us 
rewrite the above with $\delta_0 \equiv
\delta_{\mu \mu} + \delta_{\tau \tau} - 2 \delta_{\mu \tau}$, $\delta
\equiv 2 \delta_{\mu \tau}$, $\delta' \equiv \delta_{ee} - \delta_{\mu
  \mu}/2 - \delta_{\tau \tau}/2 - \delta_{\mu \tau}$, and $\delta''
\equiv \delta_{e \mu} + \delta_{e \tau}$.  
\begin{equation}
\lambda_{ij} = \left[ \begin{array} {c@{\quad}c@{\quad}c} 1 + \delta_0 +
2 \delta + 2 \delta' & \delta'' & \delta'' \\ \delta'' & \delta & 1 +
\delta_0 + \delta \\ \delta'' & 1 + \delta_0 + \delta & \delta \end{array}
\right],
\end{equation}
so that the $exact$ eigenvectors and eigenvalues are easily obtained:
\begin{equation}\label{numax}
\left[ \begin{array} {c} \nu_1 \\ \nu_2 \\ \nu_3 \end{array} \right] =
\left[ \begin{array} {c@{\quad}c@{\quad}c} \cos \theta & \sin \theta/\sqrt 2
& \sin \theta/\sqrt 2 \\ -\sin \theta & \cos \theta/\sqrt 2 & \cos \theta/
\sqrt 2 \\ 0 & -1/\sqrt 2 & 1/\sqrt 2 \end{array} \right] \left[
\begin{array} {c} \nu_e \\ \nu_\mu \\ \nu_\tau \end{array} \right],
\end{equation}
and
\begin{eqnarray}
\lambda_1 &=& 1 + \delta_0 + 2 \delta + \delta' - \sqrt{\delta'^2 + 2
\delta''^2}, \\
\lambda_2 &=& 1 + \delta_0 + 2 \delta + \delta' + \sqrt{\delta'^2 + 2
\delta''^2}, \\
\lambda_3 &=& -1 - \delta_0.
\end{eqnarray}
leading to
\begin{equation}
\sin^2 2 \theta_{atm} = 1, ~~~ \tan^2 \theta_{sol} = {\delta''^2 \over
\delta''^2 + \delta'^2 - \delta' \sqrt {\delta'^2 + 2\delta''^2}},
\end{equation}
if $\delta' < 0$ and $|\delta''/\delta'| = 1.7$.  Assuming
$\delta''^2/\delta'^2$ and $\delta', \delta'' << \delta$, we now have
\begin{equation}
\Delta m^2_{31} \simeq \Delta m^2_{32} \simeq 4 \delta m_0^2, ~~~
\Delta m^2_{12} \simeq 4 \sqrt{\delta'^2 + 2 \delta''^2} m_0^2,
\end{equation}
where $m_0$ is the common mass of all 3 neutrinos.  This provides a
satisfactory first-order description of present neutrino-oscillation
data.  With $U_{e3} = 0$ there is no $CP$ violation in neutrino
oscillations.  However, if we assume complex $\lambda_{ij}$ then one
has one CP phase which cannot be rotated away.  Without loss of
generality, we now rewrite $\lambda_{ij}$ as
\begin{equation}
\lambda_{ij} = \left[ \begin{array} {c@{\quad}c@{\quad}c} 1 +
2 \delta + 2 \delta' & \delta'' & \delta''^* \\ \delta'' & \delta & 1
+ \delta \\ \delta''^* & 1 + \delta & \delta \end{array}
\right],
\end{equation}
where we have redefined $1 + \delta_0$ as 1, and $\delta$, $\delta'$
are real.  Assuming that $\delta'$, $Re \delta''$ and $(Im
\delta'')^2/\delta$ are all much smaller than $\delta$, one can
diagonalize this mass matrix approximately,
\begin{equation}
U_{e3} = {i Im \delta'' \over \sqrt 2 \delta}, ~~~ \delta' \to \delta' +
{(Im \delta'')^2 \over 2 \delta}, ~~~ \delta'' \to Re \delta''.
\end{equation}
obtaining that $U_{e3}$ is imaginary and $CP$ violation in neutrino
oscillations is predicted to be maximal.  There is also an interesting
relationship, i.~e.
\begin{equation}\label{ratio}
\left[ {\Delta m_{12}^2 \over \Delta m_{32}^2} \right]^2 \simeq \left[
{\delta' \over \delta} + |U_{e3}|^2 \right]^2 + \left[ {Re \delta'' \over
\delta} \right]^2.
\end{equation}
indicating that $|U_{e3}|$ is naturally of the order $|\Delta
m^2_{12}/\Delta m^2_{32}|^{1/2}$ 
$$|U_{e3}| = \O (|\Delta m^2_{12}/\Delta m^2_{32}|^{1/2})$$
in the limit $\delta',\delta''\ll \delta $.

\begin{figure}[h!]
\begin{center}
\includegraphics[height=3.5cm,width=0.55\textwidth]{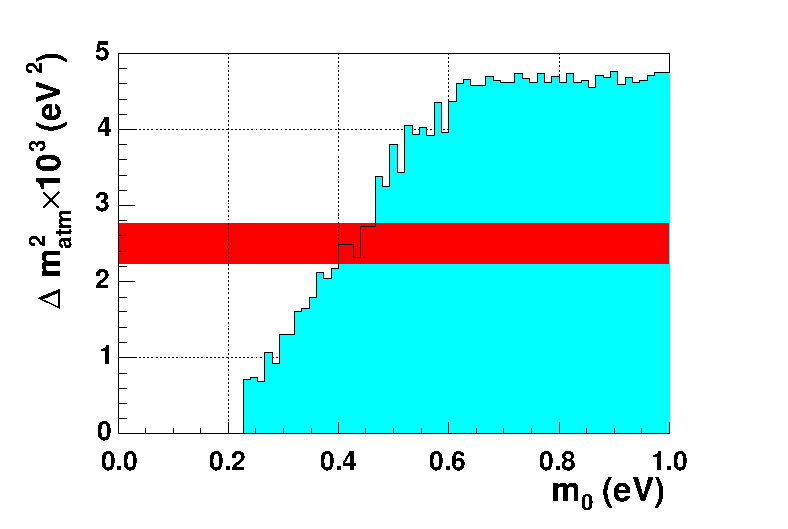}
\caption{Maximum atmospheric mass squared difference versus $m_0$
  (light shaded histogram).  The horizontal band is the 3$\sigma$
  allowed region for $\Delta m^2_{\rm atm}$ by current MINOS/T2K
  data.}
\label{mzerofig}
\end{center}
\end{figure}

In Fig.\ref{mzerofig} we plot the maximum achievable value of $\Delta
m_{\rm atm}^2$ against the overall neutrino mass scale $m_0$.  The
value of $m_0$ is subject to an upper bound given by
$(\beta\beta)_{0\nu}$
experiments\cite{Barabash:2011fn,Rodejohann:2011mu} and
cosmology~\cite{lesgourgues:2006nd,hannestad:2006zg}.
In order to get large enough neutrino mass splittings to account for
current oscillation data we need
\begin{equation}
 m_0 \gsim 0.4 \,{\rm eV} \;
\end{equation}
leaving a relatively small room for the value of $m_0$.
Note also that the mass splittings are related to the parameters
$\delta,\delta'$ and $\delta''$, and these are increasing functions of
the slepton mixings and also mass splittings. This is potentially in
conflict with the restrictions from \lfv searches which push the
spectrum toward mass degeneracy and small mixings. However one can
show that viable spectra do exist.

\section{Quarks, non-abelian discrete flavour 
symmetries and  unification}
\label{sec:quarks-non-abelian}

In the original BMV model~\cite{Babu:2002dz} as well as in the
subsequent Altarelli-Feruglio model~\cite{Altarelli:2005yp} the CKM
mixing matrix was predicted to be the identity, which provides indeed
a good first order approximation.  However, realistic quark mixings
require either renormalization
effects~\cite{hall:1985dx,borzumati:1986qx}, or suitable model
extensions, e.g~\cite{He:2006dk} and \cite{Bazzocchi:2007na}.

The largest angle in the CKM matrix is the Cabibbo angle governing the
mixing between first and second generations, about $\lambda_C\sim
0.22$.  Mixing angles between the first/third and second/third
families are about $\lambda_C^2$ and $\lambda_C^3$, respectively. This
suggests that first and second quark families belong to a doublet of a
family symmetry, instead of singlets or triplets.

A non trivial extension of the Altarelli-Feruglio model for quarks was
given in \cite{Feruglio:2007uu} where $A_4$ is extended to its double
covering $T'$. The main advantage of this group is that it contains
doublet irreps besides the triplet. This feature of $T'$ is suitable
for the quarks since the first two generations can be assigned to
doublet irreps, while the third generation belongs to a singlet.

Another group which is interesting in order to include quarks in a TBM
pattern for leptons is the permutation group of four objects $S_4$.
Lam \cite{Lam:2008sh} noted that the mininmal group for TBM is $S_4$
and Refs.~\cite{Bazzocchi:2008ej,Bazzocchi:2009pv} have studied a
model based on $S_4$.  In Ref.~\cite{Blum:2007jz} it has been shown
that the Cabibbo angle can be predicted using the dihedral flavour
symmetry $D_n$.

However in general it seems that there is not yet a simple and elegant
framework that can explain at the same time neutrinos and quarks. Most
of the models considered for quarks contain many flavon fields and
extra ad hoc abelian symmetries.


Recently there has been a lot of effort in studying the possibility to
embed the TBM \emph{ansatz} within a grand unified (GUT) framework. The most
popular unifying groups considered were $SU(5)$, Pati-Salam
$SU_c(4)\times SU_L(2)\times SU_R(2)$ and $SO(10)$. These can be
separated into two classes with respect to TBM. In the $SU(5)$
framework it is easier to obtain the TBM pattern than in the case of
Pati-Salam or $SO(10)$, since right-handed neutrinos transform
trivially under the gauge group. For some example of extensions of
$SU(5)$ with discrete flavour symmetries see for instance
Ref.~\cite{Ma:2006sk,Chen:2007afa,Altarelli:2008bg,Burrows:2009pi,Hagedorn:2010th,Cooper:2010ik,Antusch:2010es},
with Pati-Salam \cite{deMedeirosVarzielas:2006fc,Toorop:2010yh} and
\10
\cite{Cai:2006mf,Dermisek:2006dc,Morisi:2007ft,Bazzocchi:2008sp,Grimus:2008tm,King:2009tj,Dutta:2009bj,Altarelli:2010at,Blankenburg:2011vw,BhupalDev:2012nm}.

\vskip5.mm

\noindent
{\bf $SU(5)$ models}\\

As an illustrative example here we consider the model studied in
\cite{Ma:2006sk}.  In typical $SU(5)$ GUT models, the $\overline{5}$
contains the lepton doublet $L$ and the three (colored) right-handed
down type quarks $d^c$ for each family. Whereas the $10$ contains the
right handed charged leptons, the three right handed up type quarks
$u^c$ and the three quark doublets $Q$ for each family.  Typically in
$A_4$--based models the three lepton doublets are assigned as triplets
of $A_4$, while the right--handed charged leptons are assigned to
$A_4$ singlets. This means that we should choose
\begin{equation}
\overline{5}\sim 3,\qquad 10\sim 1,1',1''
\end{equation}
which implies the following quark assignments
\begin{equation}
d_i^c\sim 3,\qquad (u_i,d_i)\sim 1,1',1'',\qquad u_i^c\sim 1,1',1''.
\end{equation}
One assumes three Higgs doublets $\overline{5}_H\sim 3$ in the
down/lepton sector 
\begin{equation}
M_d=
\left(
\begin{array}{ccc}
h_1&0&0\\
0&h_2&0\\
0&0&h_3\\
\end{array}
\right)\cdot U_L\cdot 
\left(
\begin{array}{ccc}
v_1&0&0\\
0&v_2&0\\
0&0&v_3\\
\end{array}
\right)
\end{equation} 
where $h_i$ are Yukawa couplings and $v_i\equiv \langle
\overline{5}_{H_i}^0\rangle$~\footnote{ Since in  minimal $SU(5)$  we
have that $M_l=M_d^T$ it is clear that one can have TBM mixing in the
limit $v_1=v_2=v_3$.}.
For the up--quark sector one assumes only two Higgs doublets, namely
$5_H\sim 1',1''$, so that the up quark mass matrix has the form
\begin{equation}
M_u=
\left(
\begin{array}{ccc}
0&\mu_2&\mu_3\\
\mu_2&m_2&0\\
\mu_30&0&m_3\\
\end{array}
\right)
\end{equation}
where $m_2$ and $\mu_3$ arise from the vev of $5_H\sim 1'$ and $m_3$
and $\mu_2$ come from the vev of $5_H\sim 1''$.

In Ref.\,\cite{Ma:2006sk} it has been shown how the CKM mixings
$V_{us}$, $V_{us}$ and $V_{ub}$ as well as the CP phase can be
obtained in this model. In total the model has 10 free parameters in
the quark sector for 10 observable (6 masses, three mixing angles and
one phase).  Therefore it is clear that in principle such a model can
fit realistic quark masses and mixings.  Indeed this has been shown,
however no light can shed into the structure of quark masses and
mixings. 
Moreover we has the usual SU(5) relations $m_\tau=m_b$, $m_\mu=m_s$
and $m_e=m_d$ at the GUT scale. While the first relation is in good
agreement with data, the last two are not. To decouple the charged
lepton sector from the down quark sector the usual strategy is to use
bigger $SU(5)$ representations in the scalar sector, like the $45$. In
the simplest scenario the gauge coupling unification has not been
considered.

\vskip5.mm

\noindent
{\bf $SO(10)$ models}\\

The situation in $SO(10)$ with type-I seesaw is much more complicated
than in $SU(5)$.  Indeed the main problem is that in $SO(10)$ TBM
requires to distinguish the Dirac neutrino Yukawa coupling from that
of the up-quark. In particular the former must be either proportional
to the identity or given by eq.(\ref{mTBM}) in the basis where charged
leptons are diagonal. In contrast the up quarks must be strongly
hierarchical, namely
\begin{equation}\label{hier}
M_{u}\sim
\left(
\begin{array}{ccc}
\times&\times&\times\\
\times&\times&\times\\
\times&\times&1\\
\end{array}
\right),
\qquad
m_{D}\sim
\left(
\begin{array}{ccc}
1&0&0\\
0&1&0\\
0&0&1\\
\end{array}
\right),
\end{equation} 
where $\times$ indicate small entries much smaller then one.  We
consider for simplicity the case where the Dirac neutrino mass matrix
is proportional to the identity. As already mentioned, in $SO(10)$ all
matter fields (including right-handed neutrinos) belong to only one
multiplet, the spinorial $16$.  In renormalizable $SO(10)$ models only
three types of Yukawa contractions are possible, namely $16\cdot 16 =
10 +120+126$.  Two of them $Y_{10}$ and $Y_{\overline{126}}$ are
symmetric and $Y_{120}$ is antisymmetric.  It is well known that their
contributions to up quark and Dirac neutrino mass matrices are given as
\begin{equation}\label{unu}
M^u=M^\nu = Y_{10} v_5,\quad M^u= Y_{\overline{126}} r_5, \,M^\nu= -3 Y_{\overline{126}} r_5,\qquad
M^u= Y_{120} u_{45},\,M^\nu= Y_{120} u_5
\end{equation}
where $v_\alpha,u_\alpha,r_\alpha$ are the vevs of $10$, $120$ and
$\overline{126}$ respectively, and the lower indexes $\alpha=5, \,45$
are $SU(5)$ components.  Note that the $10$ gives equal contributions
to the Dirac and up-quark mass matrices, while the $ \overline{126} $
gives contributions to the Dirac and up-quark mass matrices that are
proportional to each other.  Therefore if we take hierarchical
$Y_{10}$ and $Y_{\overline{126}}$ the Dirac neutrino mass matrix will
be hierarchical, in contrast with TBM requirement given in
eq.~(\ref{hier}).  The $120$ gives different contributions to the up
quarks and Dirac neutrino mass matrices, proportional to $u_{45}$ and
$u_5$, respectively.  Therefore one can in principle distinguish
between Dirac neutrino and up quarks by means of the $120$.  However
the coupling $Y_{120}$ is antisymmetric giving two degenerate
eigenvalues and zero determinant. Hence it is not possible to obtain a
hierarchical Yukawa coupling with $120$.  We conclude that it is not
possible to obtain the TBM mixing pattern eq.(\ref{hier}) within a
simple renormalizable $SO(10)$ framework with type-I seesaw mechanism
as described above.
This problem has been circumvented in two ways:  
\begin{itemize}
\item assuming type-II dominant with respect to type-I seesaw;
\item introducing non-renormalizable operators.
\end{itemize}

The idea of type-II dominance was suggested in
Ref.~\cite{ioannisian:1994nx} in the context of an \10 model with
quasi-degenerate neutrinos. A similar scenario has been used in
Ref.~\cite{Dutta:2009bj} to accommodate the TBM mixing pattern.
The idea is that in \10 the type-II seesaw arises from the coupling
with $\overline{126}$ scalar and is proportional to its $15$ component
under $SU(5)$.  It is well known that the $15$ does not give
contributions to the Dirac fermion masses, see eqs.(\ref{unu}).
Therefore assuming type-II dominance neutrino and charged fermion
Yukawa couplings become unrelated and we can easily obtain TBM mixing
pattern.

In fact we can take the Yukawa coupling $Y_{126}$ with TBM form given
in eq.(\ref{mTBM}), while Dirac neutrino mass matrix and up quark mass
matrix can be taken hierarchical. This is no longer a problem since
type-I seesaw contribution is assumed to be negligible with respect to
type-II contribution, giving only deviations from the TBM pattern that
can generate a sizable $\theta_{13}$ angle.

The second possibility to address the problem of having a TBM mixing
patter within the \10 type-I seesaw mechanism is to use
non-renormalizable operators.  For instance one can use the dimension
five operator\,\footnote{This operator can be obtained by integrating
  out a couple $16_\chi-\overline{16}_\chi$ with renormalizable
  couplings $16 \,16_\chi\,120_H$, $16\,\overline{16}_\chi\, 45_H$.  }
\begin{equation}\label{dim5}
h\,16\,16\,120_H\,45_H.
\end{equation}
This gives a contribution to the up-quark mass matrix but not to the
Dirac neutrino~\cite{Ross:2002fb,King:2003rf,Morisi:2007ft}. 
This can be seen in more details as follows.  The $45_H$
can take vev in its singlet $1_{SU(5)}$ component called
$X$-direction\footnote{This is the extra $U(1)$ contained in
  $SO(10)\supset SU(5)\times U_X(1)$.}  or along the adjoint
$24_{SU(5)}$ component, that is the hypercharge $Y$-direction (see for
instance \cite{Anderson:1993fe}).  We indicate their vevs as
\begin{equation}\label{v124}
b_1=\langle 1_{SU(5)} \rangle,\qquad b_{24}=\langle 24_{SU(5)} \rangle.
\end{equation}
The $SU(5)$ components of the $120_H$ of $SO(10)$ that contain $SU(2)$
doublet (giving rise to the Dirac masses terms for the fermions) are
the $45_{SU(5)}$, $\overline{45}_{SU(5)}$, $\overline{5}_{SU(5)}$ and
$5_{SU(5)}$ representations. Denoting their vevs as
\begin{equation}\label{ai}
a_5=\langle 5_{SU(5)} \rangle,\qquad 
a_{\bar{5}}=\langle \overline{5}_{SU(5)} \rangle,\qquad 
a_{45}=\langle 45_{SU(5)} \rangle,\qquad 
a_{\bar{45}}=\langle \overline{45}_{SU(5)} \rangle,
\end{equation}
we find that
\begin{eqnarray}
M_u &=&h\, a_{45}(b_1-4b_{24})-h^T\, a_{45}(b_1+b_{24}),\\
M_\nu&=& 5\, h\, a_{5}b_1 - h^T\, a_{5}(-3b_1-3b_{24}),\\
M_d&=&h (a_{\overline{5}}+a_{\overline{45}})(-3b_1+2b_{24})
-h^T (a_{\overline{5}}+a_{\overline{45}})(b_1+b_{24}) ,\\
M_e^T&=&h (a_{\overline{5}}-3 a_{\overline{45}})(-3b_1-3b_{24})
-h^T (a_{\overline{5}}- 3 a_{\overline{45}})(b_1+6b_{24}) .
\end{eqnarray} 
where $h$ is an arbitrary matrix. Note that if we set $b_{24} =0$,
$M_u$ is proportional to $h-h^T$ which is antisymmetric. In contrast,
if we set $b_1=0$ and $b_{24}\ne 0$ then $M_u\propto 4 h+h^T$ which is
an arbitrary matrix.  When $a_5=0$ the operator $16\,16\,120_H\,45_H$
contributes to $Y_u$, $Y_d$, $Y_e$ but not to $Y_\nu$, so we can
distinguish between Dirac neutrino and up-quark mass matrices.

\vskip5.mm

Recent observations of a relatively large reactor
angle~\cite{An:2012eh,Abe:2011fz,Ahn:2012nd} pose the questions as to
whether the TBM pattern is indeed a good starting point to describe
neutrino flavour mixing.  The difficulty encountered in \10 based GUT
models suggests us to discard the TBM \emph{ansatz} as starting
point. However discrete non-Abelian flavour symmetries appear to be
better candidates for a family symmetry.  Here to give an explicit
example of such possibility, briefly presenting a model given in
Ref.\,\cite{Dermisek:1999vy} based on $SO(10)\times D_3$~\footnote{An
  extra $U(1)$ family symmetry is also required.}. The group $D_3$
(isomorphic to $S_3$--the group of permutations of three objects)
contains three irreducible representations, one symmetric singlet, one
antisymmetric singlet and one doublet. The third generation is
assigned to the antisymmetric singlet, $16_3\sim 1_A$ while the first
two families are assigned to a doublet of $D_3$. The only
renormalizable coupling is for the third generation, with first and
second generation masses arising only from next to leading order
contributions. Extra messenger fields are introduced in order to make
the Lagrangian renormalizable. All charged fermion Yukawa couplings
$Y_u$, $Y_d$, $Y_\ell$ and $Y_\nu$ have Fritzsch texture
\cite{fritzsch:1979zq}, namely
\begin{equation}\label{hier2}
Y_f\propto \left(
\begin{array}{ccc}
0&\times&0\\
\times&\times&\times\\
0&\times&1\\
\end{array}
\right).
\end{equation} 
Such a model is that it contains only seven real free parameters plus
three complex phases for thirteen observable charged fermion masses
and mixing angles. While in the neutrino sector there are other four
free parameters, giving in total 14 free parameters. Hence this model
is highly predictive.

\section{Dark matter and flavour symmetry}
\label{sec:dark-matter-flavour}

Deciphering the nature of Dark Matter (DM) constitutes one of the main
challenges in cosmology since decades.  Some recent direct and
indirect DM detection experiments have given tantalizing hints in
favour of a light WIMP-like DM
particle~\cite{Aalseth:2011wp,Ahmed:2010wy,Angloher:2011uu,Bernabei:2008yi,Hooper:2010mq}
feeding the hopes of an imminent detection.

An interesting idea investigated recently has been to link neutrino
mass generation to dark matter, two seemingly unrelated problems, into
a single framework. This is not only theoretically appealing, but also
may bring us new insights on both issues.

Among the requirements a viable DM candidate must obey, stability has
traditionally been ensured by imposing of a stabilizing parity in an
\textit{ad hoc} way. It would be clearly appealing to obtain stability
in a theoretically natural way.  This has motivated attempts such as
gauged $U(1)_{B-L}$~\cite{Hambye:2010zb}, gauged discrete
symmetries~\cite{Batell:2010bp} as well as the recently proposed
discrete dark matter mechanism
(DDM)~\cite{Hirsch:2010ru,Meloni:2011cc,Boucenna:2011tj,Meloni:2010sk},
where stability arises as a remnant of a suitable flavour
symmetry~\footnote{For other flavour models with DM candidates see
  \cite{Meloni:2011cc,Kajiyama:2011gu,Kajiyama:2011fx,Daikoku:2011mq,Kajiyama:2010sb,Haba:2010ag}}.

The interplay between decaying dark matter and non-Abelian discrete
flavour symmetries has been considered in a number of papers; for
instance, in \cite{Haba:2010ag,Kajiyama:2010sb,Daikoku:2010ew}
non-Abelian discrete symmetries prohibit operators that may induce too
fast dark matter decay; in \cite{Adulpravitchai:2011ei} a non-Abelian
discrete symmetry (not a flavour symmetry) has been used to stabilize
the scalar DM candidate (similar to what has been discussed in the
inert scalar models \cite{LopezHonorez:2006gr}).
However, these models differ substantially from what was proposed in
\cite{Hirsch:2010ru}.

Here we describe the original suggestion.  Consider the group of the
even permutations of four objects $A_4$.  It has one triplet and three
singlet irreducible representations, denoted ${\bf 3}$ and ${\bf
  1,1',1''}$ respectively.  $A_4$ can be broken spontaneously to one
of its $Z_2$ subgroups. Two of the components of any $A_4$ triplet are
odd under such a parity, while the $A_4$ singlet representation is
even. This residual $Z_2$ parity can be used to stabilize the DM
which, in this case, must belong to an $A_4$ triplet representation,
taken as an $SU(2)_L$ scalar Higgs doublet, $\eta\sim {\bf 3}$
\cite{Hirsch:2010ru,Meloni:2011cc,Boucenna:2011tj,Meloni:2010sk}.
Assuming that the lepton doublets $L_i$ are singlets of $A_4$ while
right-handed neutrinos transform as $A_4$ triplets $N\sim{\bf 3}$, the
contraction rules imply that $\eta$ couples only to Higgses and heavy
right-handed neutrinos $\overline{L}_i \, N\, \tilde{\eta}$.
\begin{figure}[h!]
\begin{center}
\includegraphics[width=0.33\textwidth]{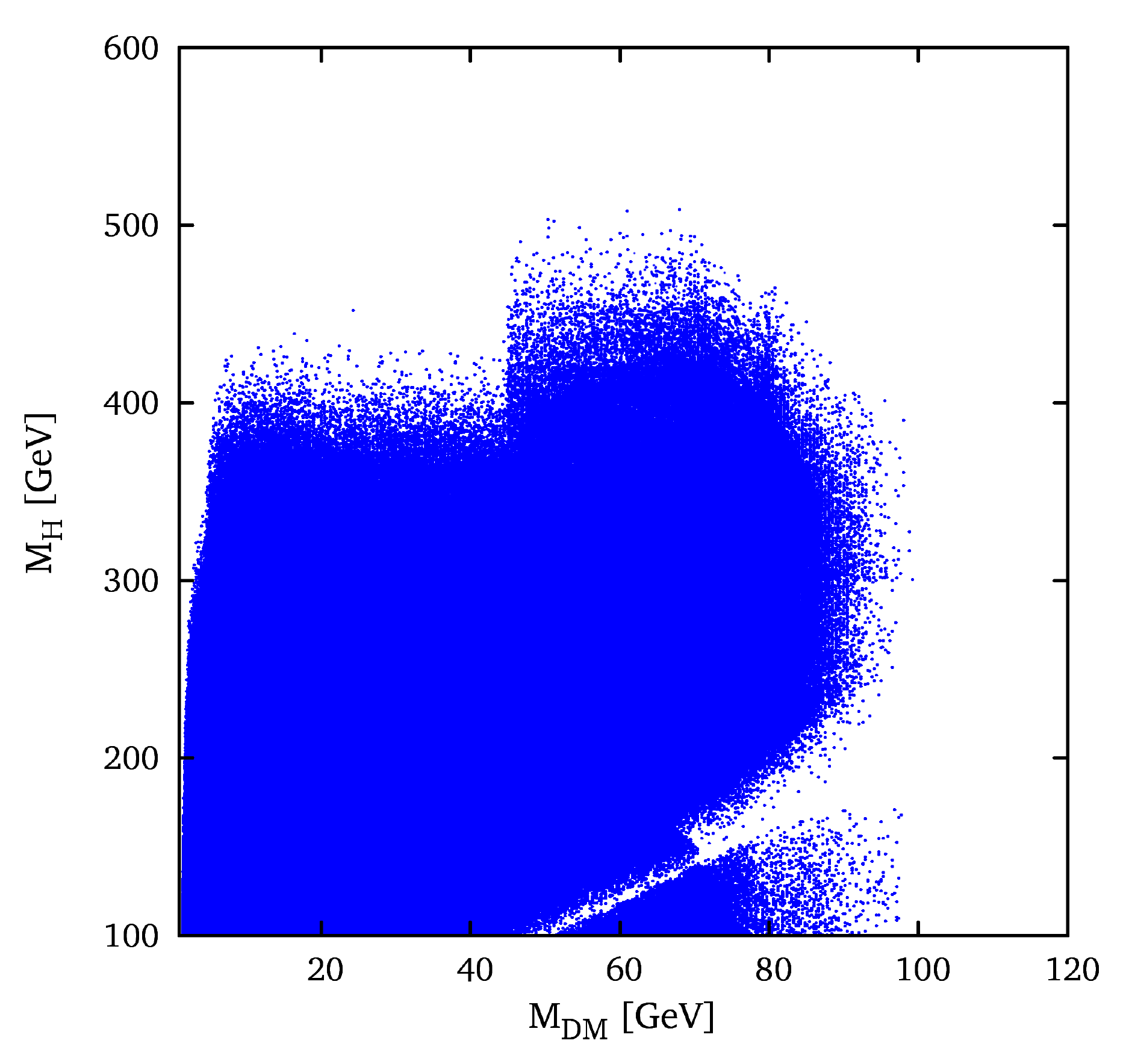}
\includegraphics[width=0.33\textwidth]{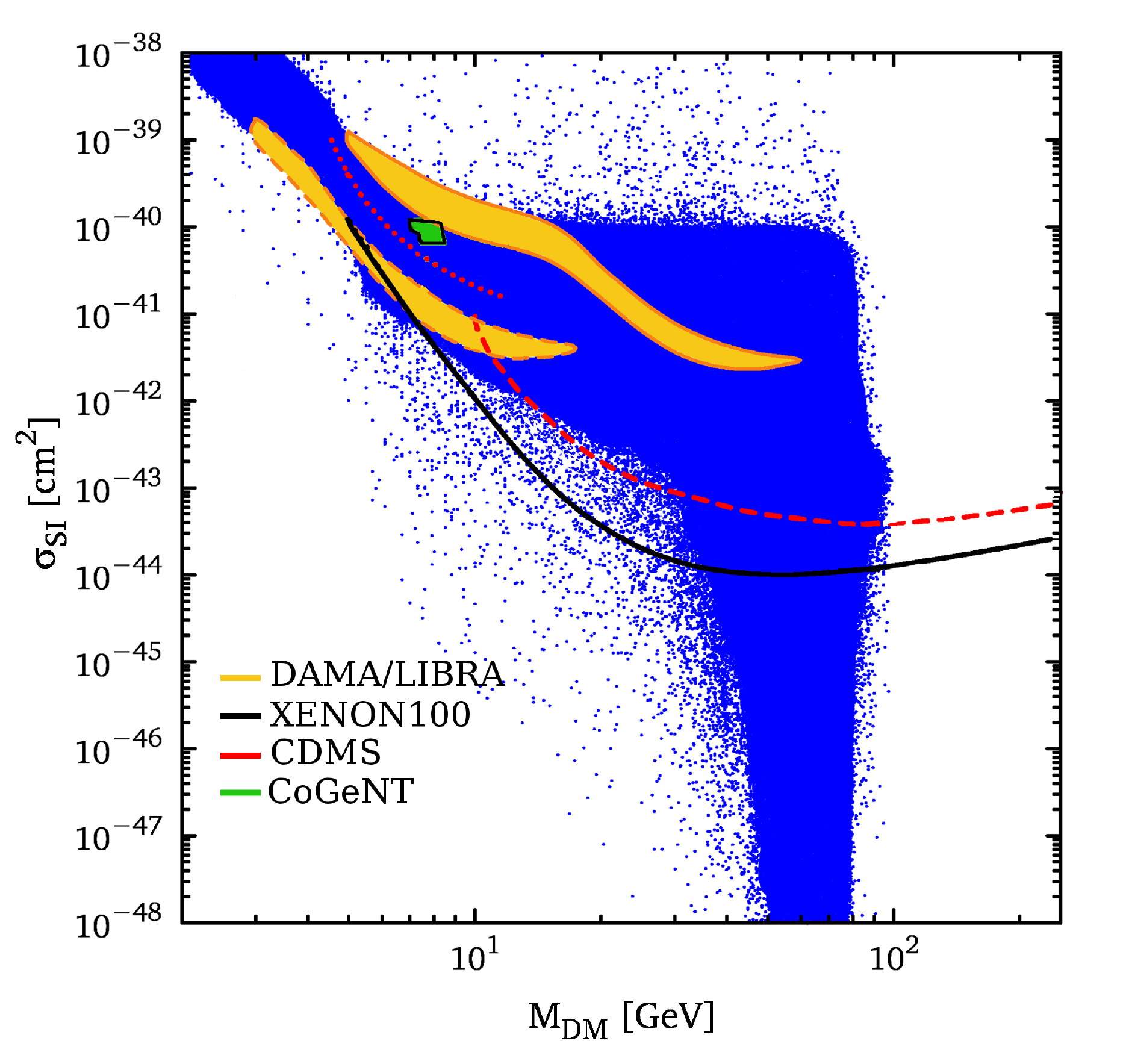}
\caption{ Left plot: Regions in ($M_{DM}$ -- $M_H$) plane (DM
  mass--lightest Higgs boson mass) allowed by collider constraints and
  leading to an adequate DM relic abundance.
  Right plot: Spin-independent DM scattering cross section off-protons
  as a function of the dark matter mass.  From
  Ref.~\cite{Boucenna:2011tj}.}
\label{omega}
\vglue -.7cm
\end{center}
\end{figure}
In this case $\eta$ and $N$ have even as well as odd-components while
$L_i$ are even so that $\overline{L}_i \, N\, \tilde{\eta}$
interaction preserves the $Z_2$ parity.  Invariance under $Z_2$
implies that $N$ components odd under $Z_2$ are not mixed with the
$Z_2$-even light neutrinos $\nu_i$. This forbids the decay of the
lightest $Z_2$-odd component of $\eta$ to light neutrinos through the
heavy right handed neutrinos, ensuring DM stability.
In this framework DM has quartic couplings with the SM Higgs doublet
as is the Higgs portal DM scenario, and has been shown to have the
correct relic density~\cite{Boucenna:2012qb}, with annihilation and
co-annihilation of DM into SM particles (fermions and bosons), see
Fig.\,\ref{omega} for the results.
Note that assigning the three left-handed leptons to a flavour-triplet
implies that the ``would-be'' DM candidate decays very fast into light
leptons, through the contraction of the triplet representations, see
general discussion in ref.~\cite{Kajiyama:2011gu}.
This problem has been considered by Eby and Framptom \cite{Eby:2011qa}
using a $T'$ flavour symmetry. While the suggested model has the merit
of incorporating quarks non-trivially, it requires an ``external''
$Z_2$ asymmetry in order to stabilize dark matter.
In fact this observation lead ref.~\cite{Lavoura:2011ry} to claim that
a successful realization of the DDM scenario requires the lepton
doublets to be in three inequivalent singlet representations of the
flavour group.
\begin{figure}[h!]
\begin{center}
 \includegraphics[height=4cm,width=7.0cm]{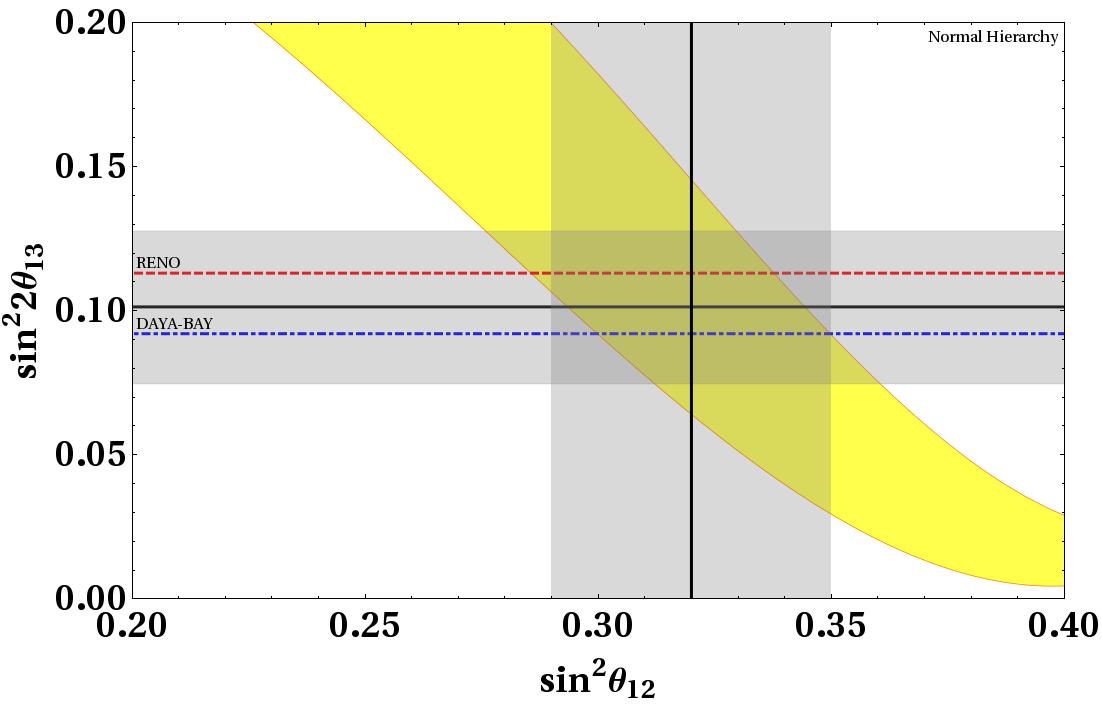}
 \caption{The shaded (yellow) curved band gives the predicted
   correlation between solar and reactor angles at two-sigma for
   normal hierarchy. The solid (black) lines give the global best fit
   from Ref.~\cite{Tortola:2012te}. The dashed lines correspond to the
   central values of the recent reactor
   measurements~\cite{An:2012eh,Abe:2011fz,Ahn:2012nd}, along with the
   corresponding two-sigma bands.}\label{fig4}
\end{center}
\vglue -.4cm \end{figure}
Recently Ref.~\cite{Boucenna:2012qb} has given an explicit example of
a model based on a $\Delta(54)$ flavour symmetry in which left-handed
leptons are assigned to non-trivial representations of the flavour
group, with a viable stable dark matter particle and a nontrivial
inclusion of quarks. In contrast to the simplest ``flavour-blind''
inert dark matter scheme~\cite{barbieri:2006dq}, such a model implies
non-trivial restrictions and/or correlations amongst the neutrino
oscillation parameters, consistent with the recent reactor angle, see
Fig.~\ref{fig4}.
Although neutrino mixing parameters in the lepton mixing matrix are
not strictly predicted, as seen in Fig.~(\ref{fig4}), there is a
non-trivial correlation between the reactor and the atmospheric angle.
While the solar angle is clearly unconstrained and can take all the
values within in the experimental limits, a nontrivial correlation
exists with the reactor mixing angle, indicated by the bands in
Fig.~\ref{fig4}. These correspond to two and $3\sigma$ regions
corresponding to the global oscillation fit in
Ref.~\cite{Tortola:2012te}.  The horizontal lines give the best global
fit value and the recent best fit values obtained in Daya--Bay and
RENO reactors~\cite{An:2012eh,Ahn:2012nd}. 

\section{Neutrinoless double beta decay}
\label{sec:neutr-double-beta}

\begin{figure}[h!]
\begin{center}
 \includegraphics[height=3.5cm,width=7.0cm]{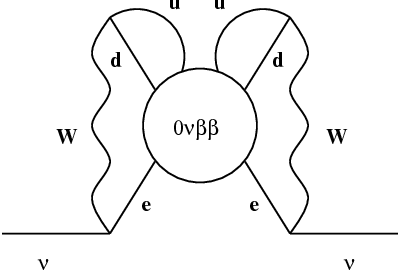}
 \caption{The observation of \znbb implies the Majorana nature of a
   neutrino~\cite{schechter:1982bd}.}\label{ref:fig-box}
\end{center}
\vglue -.7cm \end{figure}

\begin{figure}[h!]
  \begin{center}
 \includegraphics[width=75mm,height=42mm]{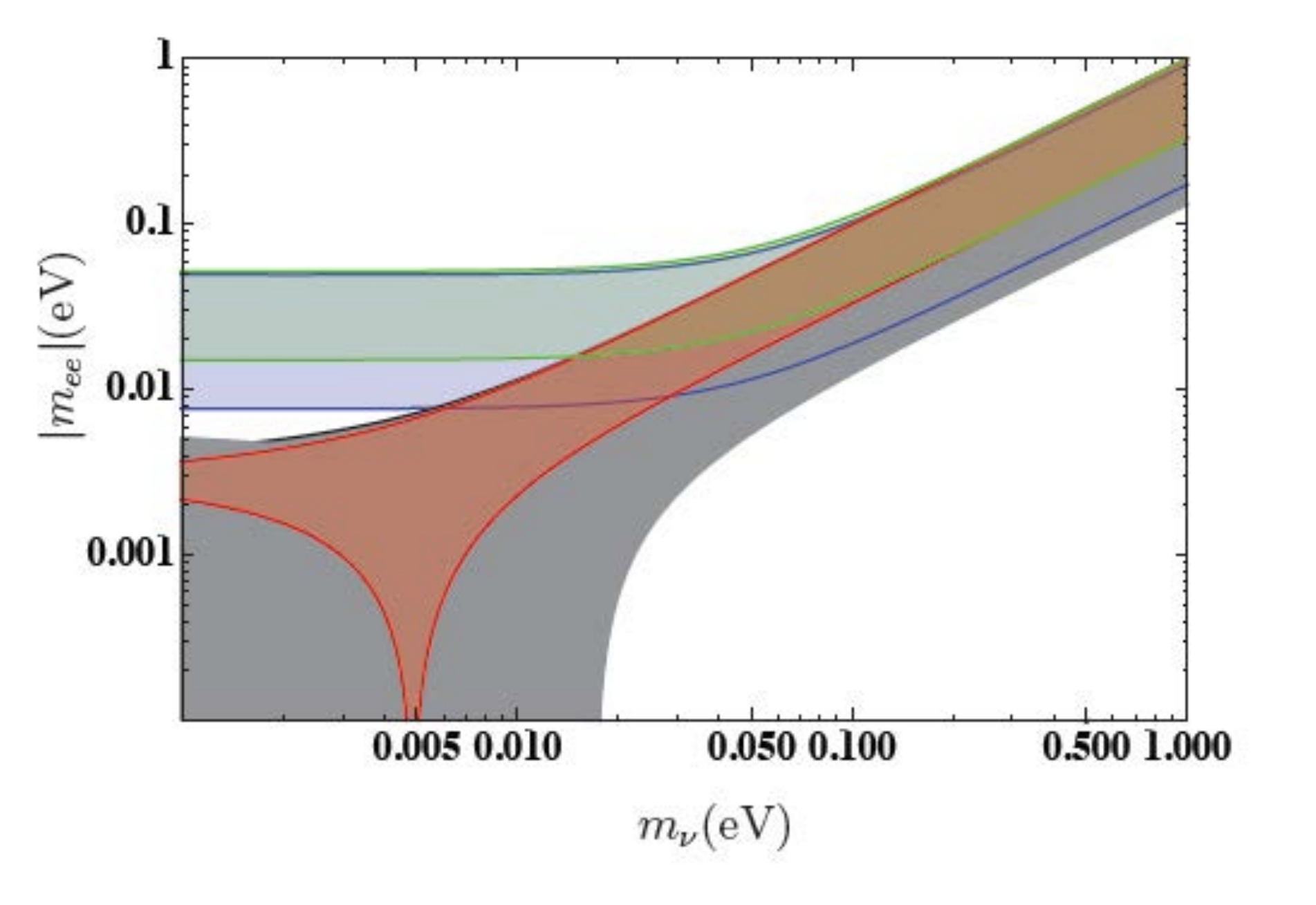}
 \end{center}
 \caption{Allowed range of $\langle |m_{ee}|\rangle$ as a function of
   the lightest neutrino mass for the TBM mixing pattern (red and
   green bands for NH and IH respectively) and for the full 3$\sigma$
   C.L. ranges of oscillation parameters from~\cite{Tortola:2012te}
   (gray and blue bands for NH and IH respectively). }
 \label{fig2}
\end{figure}

This is the process \emph{ par excellence} which allows us ways to
test the fundamental nature - Dirac or Majorana - of neutrinos in a
model-independent way, i.~e. irrespective of which mechanism generates
neutrino masses and irrespective of which mechanism induces \znbb
(neutrinoless double beta decay)~\footnote{CP and electromagnetic
  properties of
  neutrinos~\cite{schechter:1981hw,wolfenstein:1981kw,pal:1982rm,deGouvea:2002gf}
  are also sensitive to the fundamental nature of neutrinos, though
  experimental prospects are far less clear than those for \znbb.}.
The basic connection is given by the \emph{black box
theorem}~\cite{schechter:1982bd,Duerr:2011zd} illustrated in
Fig.~\ref{ref:fig-box}.
Moreover, \znbb receives a contribution from the tree-level exchange
of Majorana neutrinos, whose amplitude, illustrated in
Fig.~\ref{fig2}, is proportional to an effective mass parameter which,
in contrast to neutrino oscillations, is sensitive also to the
absolute scale of neutrino masses, which is independently tested also
in searches for tritium beta decay~\cite{Drexlin:2005zt,Masood:2007rc}
and cosmology~\cite{lesgourgues:2006nd,hannestad:2006zg}. 

In addition, this amplitude can bring complementary information on the
underlying flavour structure as revealed, say, in neutrino oscillation
searches~\footnote{Subject, of course, to nuclear matrix element
  uncertainties~\cite{faessler:2008xj}.}.

As we saw in Sec.~\ref{sec:bmv-model} the BMV model~\cite{Babu:2002dz}
implies a lower bound on the absolute neutrino mass $m_{\nu}\gsim 0.4$
eV and therefore will be tested fairly soon in \znbb searches.

On the other hand, even if neutrinos are non--degenerate, many of the
models based on non--Abelian discrete flavour symmetries are
characterized by a specific (complex) relation between neutrino mass
eigenvalues, leading to mass sum rules (MSR). Such MSR lead in most of
the cases to lower bounds for the neutrinoless double beta amplitude
parameter, depending of the type of the specific MSR.
The following types of mass relations hold:
\begin{eqnarray}
&A)&\chi\, m_2^\nu+\xi\, m_3^\nu=m_1^\nu, \label{i}\\
&B)&\frac{\chi }{m_2^\nu}+ \frac{\xi}{m_3^\nu}=\frac{1}{m_1^\nu},\label{ii}\\
&C)&\chi\,\sqrt{ m_2^\nu}+\xi\, \sqrt{m_3^\nu}=\sqrt{m_1^\nu}~. \label{iii}\\
&D)&\frac{\chi}{\sqrt{ m_2^\nu}}+\frac{\xi }{\sqrt{m_3^\nu}}=\frac{1}{\sqrt{m_1^\nu}}\,. \label{iiii}
\end{eqnarray}
Here $m_i^\nu=m_i^0$ denote neutrino mass eigenvalues, up to a
Majorana phase factor, while $\chi$ and $\xi$ are free parameters
which specify the model, taken to be positive without loss of
generality. 

We first consider the amplitude for neutrinoless double-$\beta$ decay
within a flavour-generic scheme. The effective neutrino mass parameter
$|m_{ee}|$ determining the $0\nu\beta \beta$ decay amplitude, as a
function of the lightest neutrino mass is given in Fig.~\ref{fig2}. As
is well-known, by varying the neutrino oscillation parameters in their
allowed ranges~\cite{Tortola:2012te} one obtains two types of
relatively broad bands in the $(|m_{ee}|,m_{\text{light}}^\nu)$ plane
corresponding to normal and inverse hierarchy spectra, as shown in
Fig.~\ref{fig2}.
For such ``flavour generic'' case one finds a lower bound for the
neutrinoless double-$\beta$ decay effective mass parameter $|m_{ee}|$
only in the case of inverse mass hierarchy. Indeed, thanks to the
possibility of destructive interference among the light neutrinos no
lower bound holds for the case of normal neutrino mass
hierarchy~\cite{wolfenstein:1981rk,schechter:1981hw,Valle:1982yw}.

Let us now turn to the case where ``flavoured'' case where MSR
relations like {\it (A),(B),(C)} and {\it (D)} hold. As mentioned
above these can be obtained in flavour models where the neutrino mass
matrix depends only on two independent free parameters, so that the
resulting mixing angles are fixed, as in the case of tri-bimaximal or
bimaximal mixing patterns.

For definiteness here we focus on the case of tri-bimaximal neutrino
mixing pattern. Taking into account that corrections from higher
dimensional operators and/or from the charged lepton sector can yield
$\theta_{13} \neq 0$, here we retain the TBM approximation as a useful
starting point to obtain our MSR relations. However, when evaluating a
lower bound on the effective neutrino mass parameter $|m_{ee}|$
determining the \znbb decay amplitude, we include explicitly the
effects of non-vanishing $\theta_{13}$, by taking the $3~\sigma$
oscillation parameter values determined in Ref.~\cite{Tortola:2012te}.

This way one obtains lower limits of $|m_{ee}|$ corresponding to
different integer choices of $(\chi,\xi)$ between 1 and 3 and for each
of the four MSR considered in eqs.~(\ref{i})-(\ref{iiii}), for both
normal and inverted hierarchies. These results are summarized in
Tab.~I of Ref.~\cite{Dorame:2011eb}. A large class of non-Abelian
flavour symmetry models discussed in in the literature is covered, see
for example, references
\cite{Bazzocchi:2009da,Ma:2005sha,Altarelli:2005yp,Ding:2010pc,Barry:2010zk,Altarelli:2005yx,Altarelli:2006kg,Ma:2006vq,Bazzocchi:2009pv,Bazzocchi:2007na,Bazzocchi:2007au,Honda:2008rs,
  Brahmachari:2008fn,Lin:2008aj,Chen:2009um,Ma:2009wi,Fukuyama:2010mz,Bazzocchi:2008ej,Chen:2007afa,Ding:2008rj,Chen:2009gy,
Chen:2009gy,Burrows:2010wz,Babu:2005se,He:2006dk,Morisi:2007ft,Altarelli:2008bg,Adhikary:2008au,Csaki:2008qq,Altarelli:2009kr,
  Lin:2009bw,Hagedorn:2009jy,Burrows:2009pi,Berger:2009tt,Ding:2009gh,Mitra:2009jj,delAguila:2010vg}.

\begin{figure}[h!]
  \begin{center}
  \includegraphics[width=70mm,height=45mm]{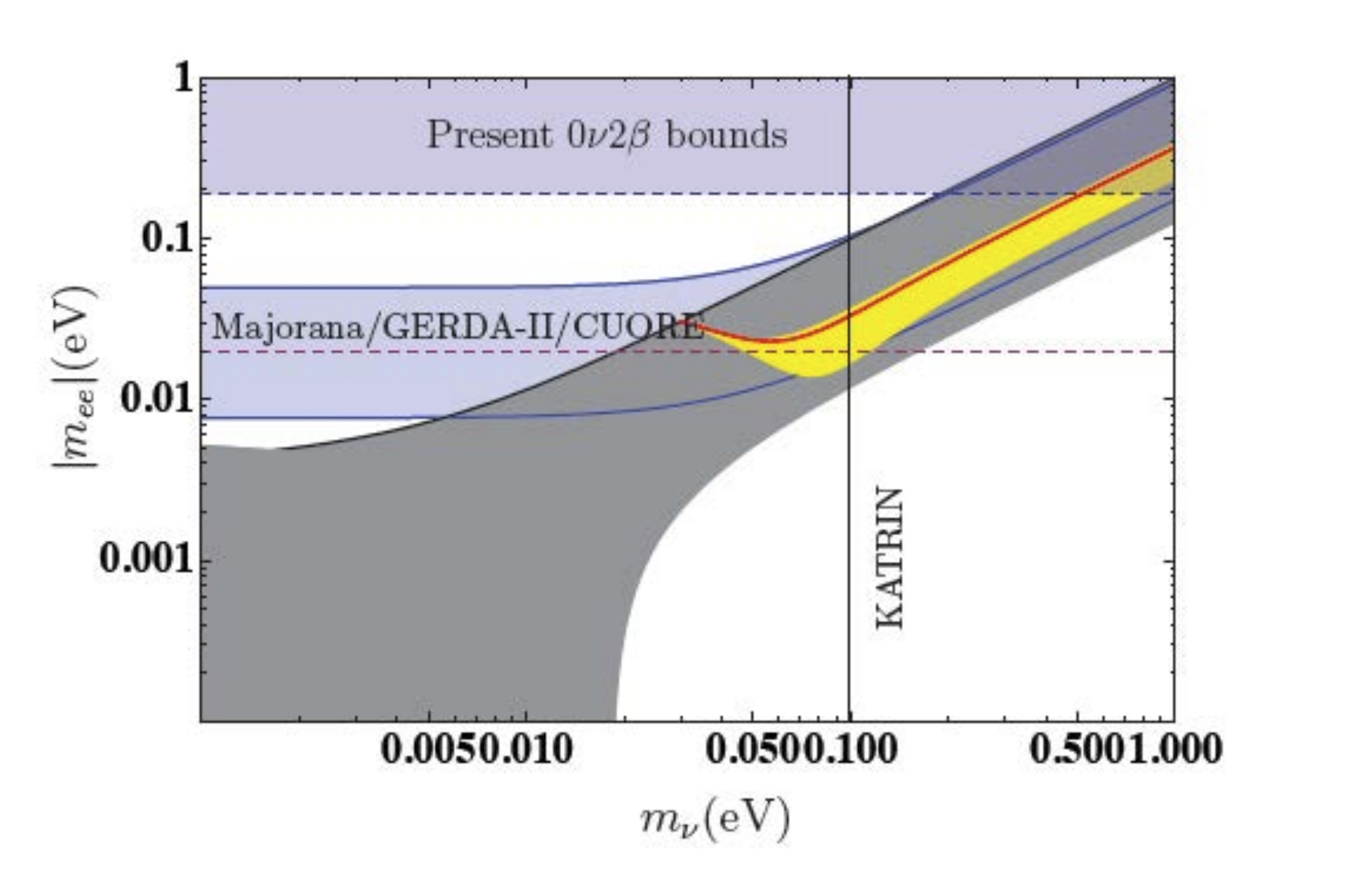}
  \includegraphics[width=70mm,height=45mm]{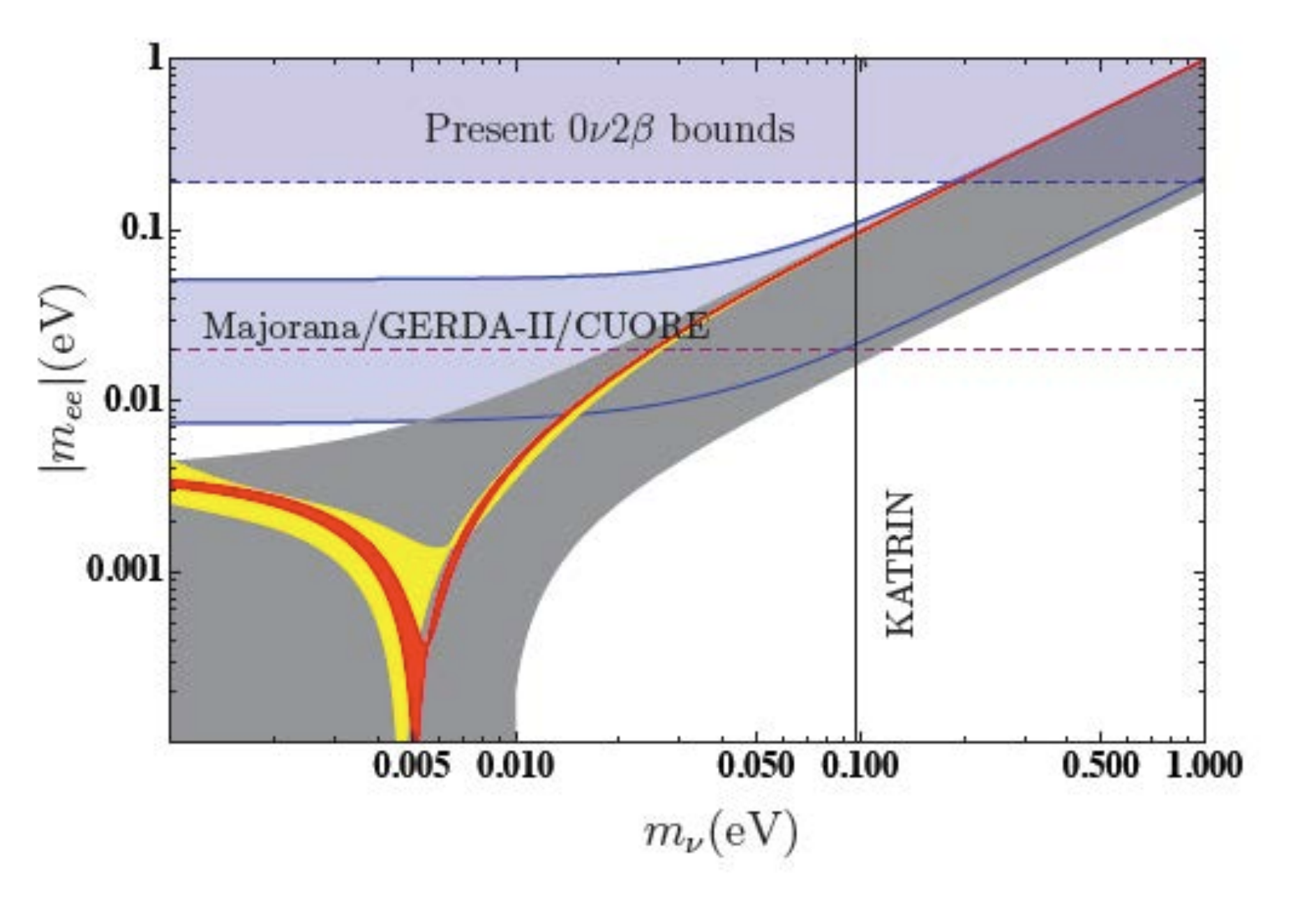}
 \end{center}
 \caption{$\langle |m_{ee}|\rangle$ as a function of the lightest
   neutrino mass corresponding to the mass sum-rule $2\sqrt{m_2} +
   \sqrt{m_3} = \sqrt{m_1}$ \cite{Hirsch:2008rp} (left) and
   $3\sqrt{m_2} + 3\sqrt{m_3} = \sqrt{m_1}$ (right).  The red bands
   correspond to the TBM mixing pattern, while the yellow bands
   correspond to the same MSR, but now varying the values of
   $\theta_{13}$, $\theta_{23}$ and $\theta_{12}$ to 3$\sigma$
   C.L. range.}
  \label{fig3}
\end{figure}

Some comments are in order. First let us consider the effect of a
possible non-zero effect of $\theta_{13}$ as indicated by recent
experiments~\cite{Abe:2011sj,Abe:2011fz} as well as global neutrino
oscillation fits~\cite{Tortola:2012te,Fogli:2011qn}.
In Fig \ref{fig3} we show the prediction for $|m_{ee}|$ as function of
$m_{light}$ obtained from the MSR $3\sqrt{m_2} + 3\sqrt{m_3} =
\sqrt{m_1}$ (right panel) and $2\sqrt{m_2} + \sqrt{m_3} = \sqrt{m_1}$
(left panel). For the red bands we assumed the TBM values of the
oscillation parameters (implying $\theta_{13}=0$) while the yellow
bands corresponds to the same MSR, but now varying the values of
$\theta_{13}$, $\theta_{23}$ and $\theta_{12}$ within their allowed
3$\sigma$ C.L. interval.
By looking at the left panel in Fig.~\ref{fig3} one sees that, indeed,
the \znbb lower bound is sensitive to the value of $\theta_{13}$.

One also finds that, as expected on general grounds, all inverse
hierarchy schemes corresponding to various choices of $(\chi,\xi)$
within sum-rules A-D have a lower bound for the parameter
$|m_{ee}|$. However, the numerical value obtained depends on the MSR
scheme, signaling that not all values within the corresponding band in
Fig.~\ref{fig2} are covered for a given flavour symmetry structure.


\section{Lepton flavour violation and flavour symmetry}
\label{sec:lept-flav-viol}

\begin{figure}[h] \centering
\vglue -.4cm 
  \includegraphics[height=3.5cm,width=.8\linewidth]{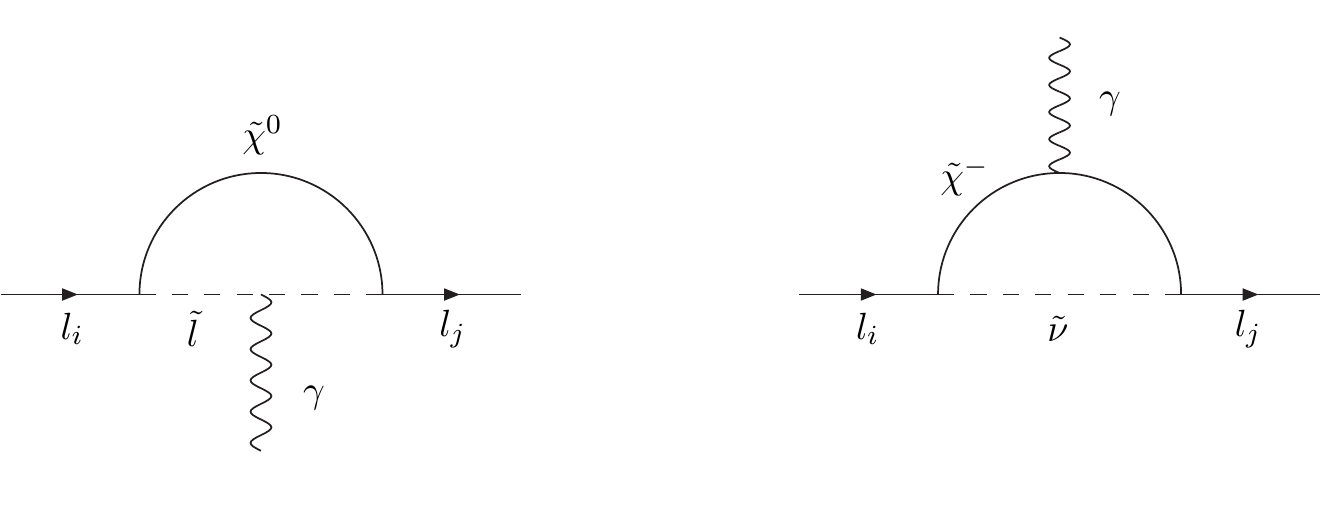}
  \caption{\label{fig:mueg1} Feynman diagrams for \(l_{i}^{-} \to
    l_{j}^{-}\gamma\) involving chargino/neutralinos and
    sneutrino/charged slepton exchange.}
\vglue -.4cm 
\end{figure}

\begin{figure}[h!]
\centering
 \includegraphics[clip,height=3.5cm,width=0.8\linewidth]{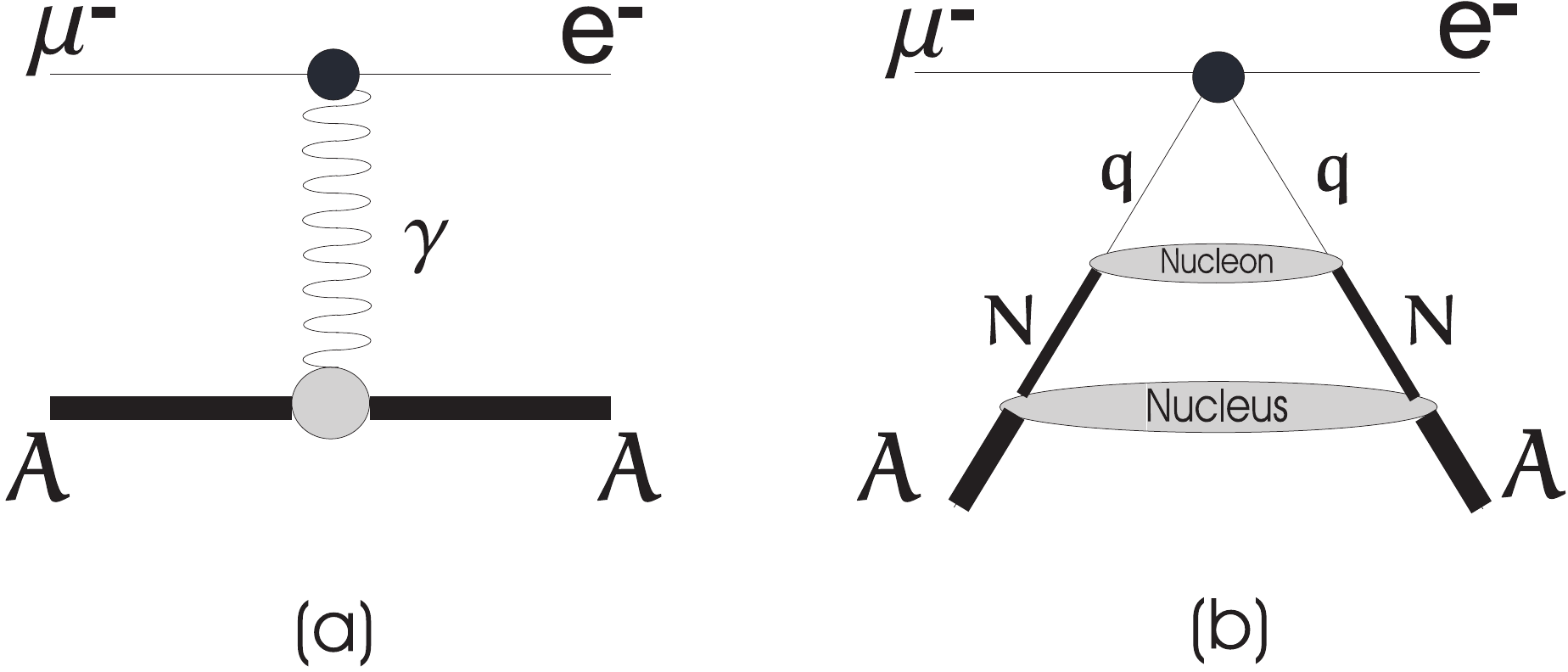}
\caption{Contributions to the nuclear $\mu^-- e^-$ conversion: (a)
  long-distance and (b) short-distance.}
     \label{fig:Diagrams}
\end{figure}

Flavour violation is required to account for the current neutrino
oscillation data. It should, however, make its appearance also in
other sectors, inducing rare processes involving charged leptons,
whose strength is not suppressed by the smallness of neutrino
masses~\cite{hall:1986dx,bernabeu:1987gr}~\footnote{Similar results
  hold also for leptonic CP
  violation~\cite{branco:1989bn,rius:1990gk}.}.
For example, in the presence of supersymmetry, the \lfv required to
account for oscillation data in high-scale seesaw schemes induces
decays such as $\mu^-\to e^-\gamma$ and flavour violating tau decays
(Fig.~\ref{fig:mueg1}) as well as nuclear $\mu^--e^-$ conversion
(Fig.~\ref{fig:Diagrams}) as a result of the exchange of
supersymmetric leptons. The existence of such loop effects,
illustrated in Fig.~\ref{fig:mueg1}, has been known for a
while~\cite{hall:1986dx,borzumati:1986qx}.
These \lfv processes can have interesting rates, which depend not only
on the seesaw mechanism, but also on the details of supersymmetry
breaking and on a possible theory of flavour. The resulting \lfv rates
will be accessible to the upcoming generation of
experiments~\cite{Kurup:2011zza,Kutschke:2011ux,Deppisch:2012vj}.

\subsection{Non-supersymmetric low-scale seesaw models and flavour}
\label{sec:non-supersymm-low}

Low energy seesaw schemes such as the inverse or linear seesaw
mechanisms generate neutrino masses from fermion messengers
(right--handed neutrinos) at the TeV
scale~\cite{mohapatra:1986bd,akhmedov:1995ip,akhmedov:1995vm,malinsky:2005bi}.
These are potentially accessible to the LHC, especially in the
presence of a new gauge boson ``portal'' associated, for example, to
left-right symmetry~\cite{Das:2012ii,AguilarSaavedra:2012fu}.

Within low-scale seesaw mechanisms \lfv and/or CP violating effects
arise at the one--loop level from the exchange of relatively light
neutral heavy leptons. Their strength is not suppressed by the
smallness of neutrino masses~\cite{hall:1986dx,bernabeu:1987gr} so the
resulting \lfv effects are potentially large even in the absence of
supersymmetry~\cite{bernabeu:1987gr,branco:1989bn,rius:1990gk}
~\cite{gonzalez-garcia:1992be,Ilakovac:1994kj} and/or extended gauge
structure~\cite{Das:2012ii,AguilarSaavedra:2012fu}~\footnote{In type-I
  seesaw schemes the processes \(l_{i}^{-} \to l_{j}^{-}\gamma\) are
  enhanced due to a breakdown of the GIM mechanism. However this is
  not enough to give large rates since the messenger scale $M_R$
  characterizing \lnv is too high.}.

\begin{figure}[!h]
\begin{center}
\includegraphics[angle=0,height=4.5cm,width=0.45\textwidth]{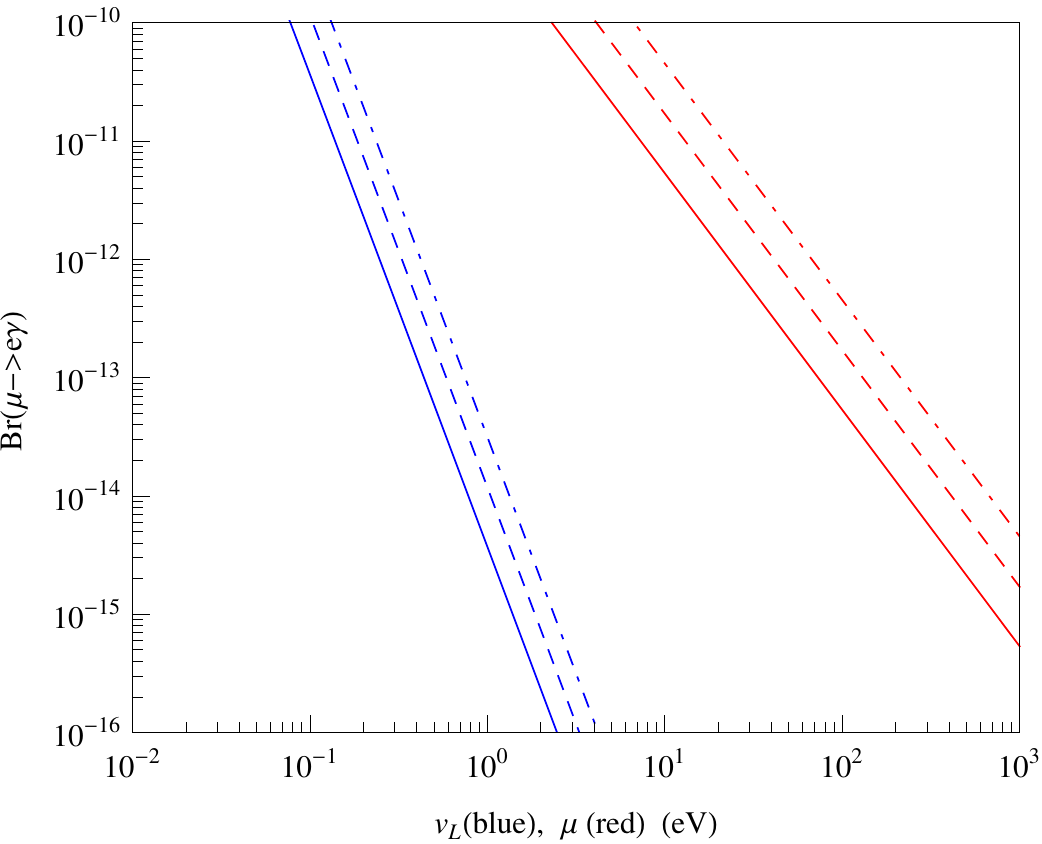}
\caption{$Br(\mu\to e \gamma)$ versus the lepton number
  violation scale: $\mu$ for the inverse seesaw (red color), and $v_L$
  for the linear seesaw (blue color). Here $M$ is fixed as $M=100 \,
  GeV$ (continuous line), $M=200\, GeV$ (dashed line) and $M=1000\,
  GeV$ (dot-dashed line).
}
\label{fig:MEGBR}
\end{center}
\end{figure}

In the inverse and linear seesaw models proposed in
\cite{Hirsch:2009mx}, the neutrino mass matrix is a $9\times 9$
symmetric matrix.  It is diagonalized by a unitary matrix
$U_{\alpha\beta}$, $\alpha,~\beta=1...9$, leading to three light
Majorana eigenstates $\nu_{i}$ with $i=1,2,3$ and six heavy ones $N_j$
with $j=4,..,9$.
The effective charged current weak interaction is characterized by a
rectangular lepton mixing matrix $K_{i\alpha}$~\cite{schechter:1980gr},
\begin{equation}
\mathcal{L}_{CC}=\frac{g}{\sqrt{2}}K_{i\alpha}\overline{L}_i\gamma_\mu (1+\gamma_5) N_\alpha \, 
W^{\mu},
\end{equation}
where $i=1,2,3$ denote the left-handed charged leptons and $\alpha$
the neutrals.  The contribution to the decay \(l_i \to l_j \gamma\)
arises at one loop from the exchanges of the six heavy right-handed
Majorana neutrinos $N_j$ which couple sub-dominantly to the charged
leptons.
The well-known one--loop contribution to this branching ratio is
given by \cite{Ilakovac:1994kj} 
\begin{equation}
Br(l_i\to l_j \gamma)= \frac{\alpha^3s_W^2}{256 \pi^2}\frac{m_{l_i}^5}{M_W^4}
\frac{1}{\Gamma_{l_{i}}}|G_{ij}|^2
\end{equation}
where
\begin{equation}\label{def:G}
\begin{array}{l}
G_{ij}=\sum_{k=4}^9 K^*_{ik} K_{jk} G_\gamma\left(\frac{m^2_{N_k}}{M_W^2}\right)\\
G_\gamma(x)=-\frac{2 x^3+5 x^2 -x}{4 (1-x^3)}-\frac{3 x^3}{2(1-x)^4}\ln x
\end{array}
\end{equation}
We note that, thanks to the admixture of the TeV neutral leptons in
the charged current weak interaction, this branching ratio can be
sizeable~\cite{bernabeu:1987gr}. Similar results hold for a class of
LFV processes, including nuclear mu-e
conversion~\cite{Deppisch:2005zm} whose expected rates are strongly
correlated to those of \(\mu \to e \gamma\), see Fig.~\ref{fig:mueg3}.
As an example we now consider the low-scale seesaw TBM model given in
\cite{Hirsch:2009mx}.  The simplicity of their mass matrices, which
are expressed in terms of very few parameters, makes such a models
especially restrictive and this has an impact in the expected pattern
of LFV decays.
In contrast to the general case considered in
\cite{Deppisch:2004fa,Deppisch:2005zm}, in \cite{Hirsch:2009mx} one
can easily display the dependence of the $\mu \to e\gamma$ branching
ratio on the new physics scale represented by the parameters $M \sim$
TeV and the parameters $\mu$ or $v_L$ characterizing the low-scale
violation of lepton number. This is illustrated in
Fig.~\ref{fig:MEGBR}.

Note also that, in contrast to a generic inverse or linear seesaw
model, such a  $A_4$--based models the structure of the matrix $G_{ij}$
is completely fixed, and this leads to predictions for ratios of \lfv
branching ratios such as
\begin{equation}
\frac{Br(\tau\to e \gamma)}{Br(\mu\to e \gamma)}=
\left(\frac{m_\tau}{m_\mu}\right)^5\frac{\Gamma_\mu}{\Gamma_\tau}\approx 0.18,
\end{equation}
for both linear and inverse seesaw and
$Br(\tau\to\mu\gamma)/Br(\tau\to e \gamma)$.

\subsection{Supersymmetric high-scale seesaw and flavour symmetry}
\label{sec:supersymm-sees-with}

In the presence of supersymmetry, the \lfv observed in neutrino
oscillations induces decays like $\mu^-\to e^-\gamma$, flavour
violating tau decays as well as nuclear $\mu^--e^-$ conversion
(Fig.~\ref{fig:Diagrams}) through the exchange of supersymmetric
leptons, as discussed for example,
in~\cite{Antusch:2006vw,Calibbi:2006nq,Joaquim:2006uz}
and~\cite{hirsch:2008dy,esteves:2009vg}.

Instead of considering a generic ``flavour-blind''supersymmetric
high-scale seesaw scheme here we consider, as an example, the BMV
model already introduced in Sec.~\ref{sec:bmv-model}.
A detailed numerical analysis of \lfv rates has been performed in
Ref.~\cite{Hirsch:2003dr}, using constraints from neutrino oscillation
data and confronting with \lfv searches~\cite{nakamura2010review}.
The allowed parameter space is determined by a random search through
the multi-dimensional parameter space, keeping all supersymmetric
masses real and in the range 100 GeV to 1000 GeV.
The strongest bounds on \lfv come from $\ell_j \to \ell_i \gamma$. 

The allowed range for the charged slepton parameters is quite
restricted. The spectra fall into two different groups.  The normal
hierarchy having two low mass sleptons ($\sim 150$ GeV) and one heavy
(above $\sim$ 500 GeV), and the inverted hierarchy case having two
heavy sleptons and one light. In both cases at least one slepton mass
lies below about 200 GeV, detectable at the LHC.  Most points fall
into the case of normal hierarchy, which often corresponds to a normal
hierarchy for the neutrinos as well.  The typical case has one small
and two large mixing angles.
Evidently the small mixing angle is needed to suppress the decay $\mu
\to e \gamma$.  Also the degeneracy of two of the sleptons
helps to minimize the LFV. As a rule of thumb there is at least one
pair of sleptons with a mass splitting of less than 40 GeV.

\begin{figure}[h!]
\begin{center}
   \includegraphics[width=0.30\textwidth]{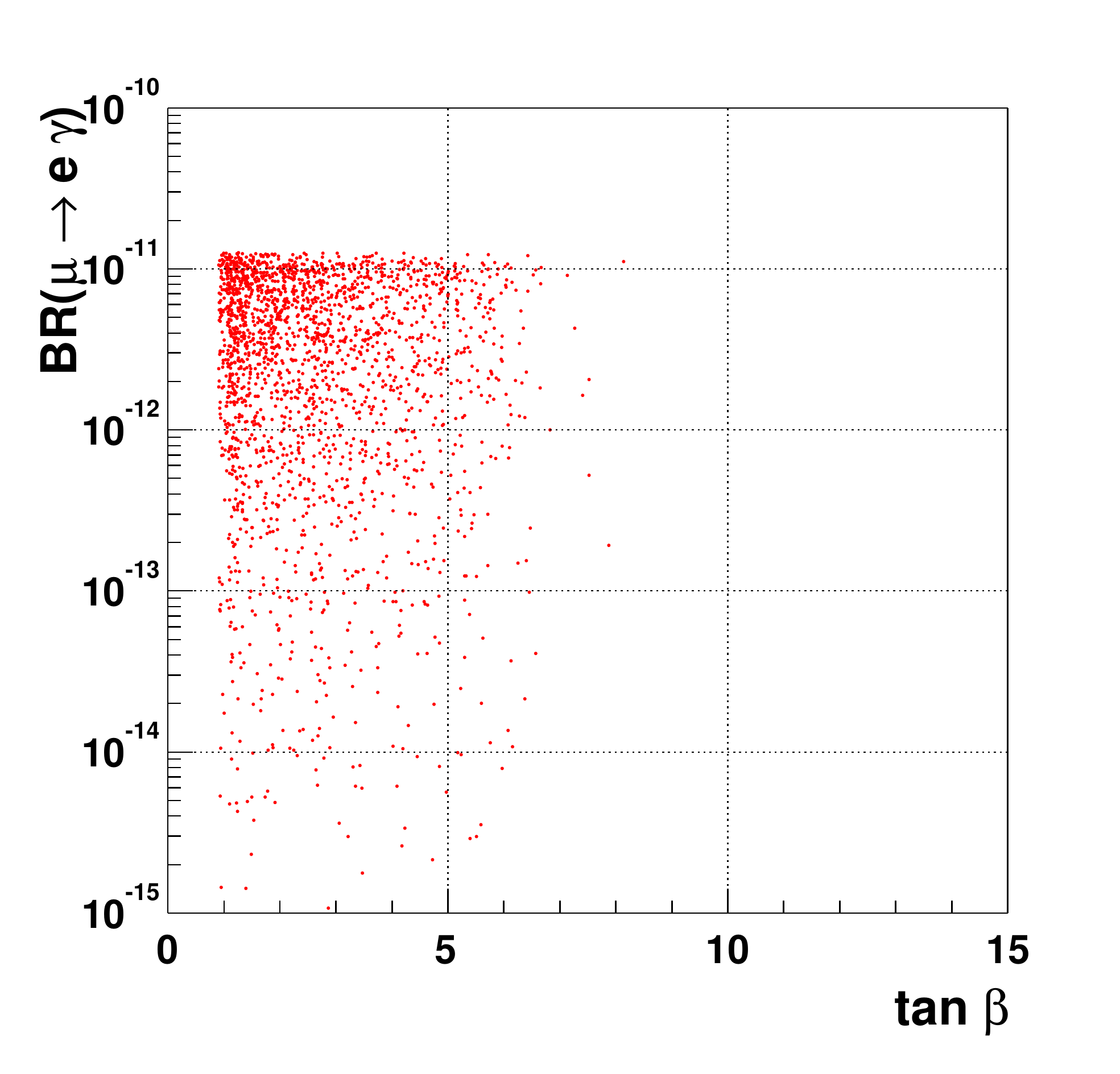}
   \includegraphics[width=0.30\textwidth]{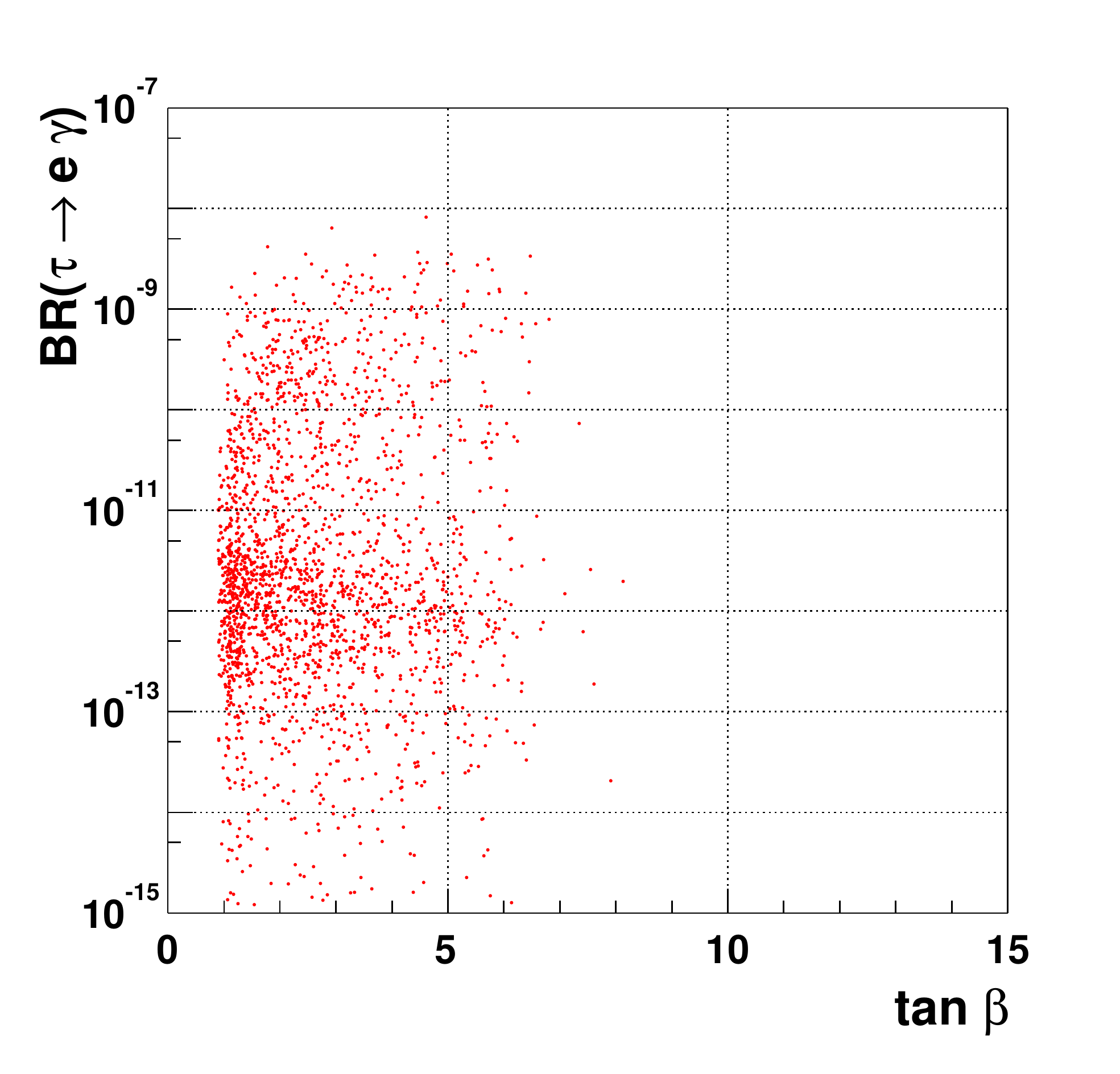}
   \includegraphics[width=0.30\textwidth]{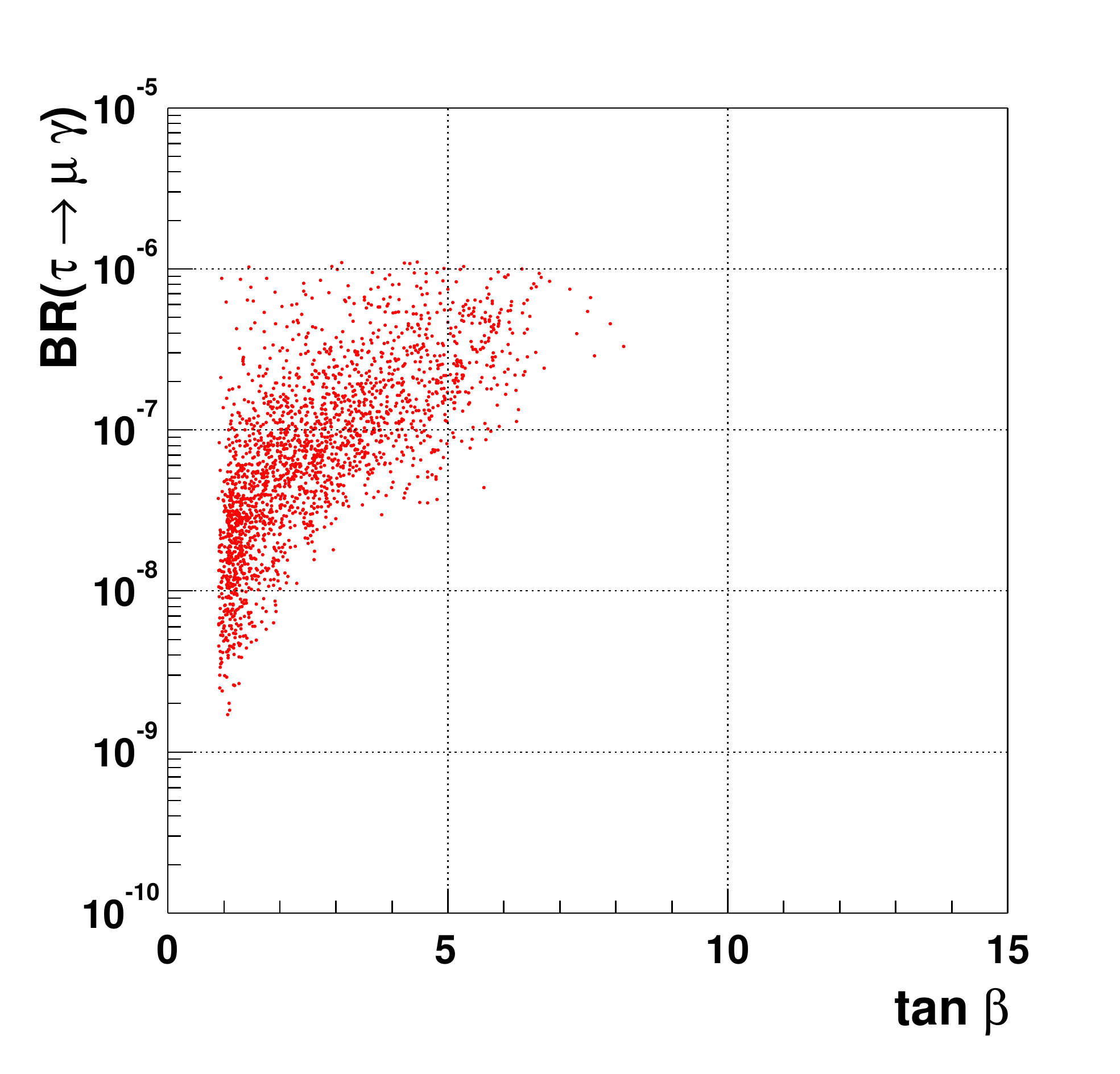}
\caption{The predictions for the branching ratios for the processes 
$\ell_i \to \ell_j \gamma$ as a function of $\tan(\beta)$. 
}
\label{brfig}
\end{center}
\end{figure}

An important outcome of this study is the prediction for the charged
lepton decays $\ell_i \to \ell_j \gamma$. As seen in Fig.\ref{brfig} a
lower bound of $10^{-9}$ for BR$(\tau \to \mu \gamma)$ is found. The
is within reach of BaBar and Belle searches. Also, BR$(\mu \to e
\gamma)$ is constrained to be larger than about $10^{-15}$ and
therefore stands good chance of being observed in the
future~\cite{Kutschke:2011ux}.
The value of $\tan(\beta)$, also plotted in Fig.\ref{brfig}, is
constrained to be small. For large $\tan(\beta)$ the RGE effect
destroys the agreement with the solar data.  The numerical value of
$\delta_\tau$ can not be much bigger than the solar mass scale.  A
rough estimate gives $\delta_\tau \stackrel{<}{\sim} 5 \times
10^{-4}$, corresponding to the bound $\tan(\beta)<10$. This agrees
with the precise bound found in Fig.\ref{brfig}. For small values of
$\tan(\beta)$ the threshold corrections dominate and the strongest
constraint comes from the bound on $BR(\mu \to e \gamma$).

\subsection{Low-scale seesaw and \lfv}
\label{sec:low-scale-seesaw-1}

\begin{figure}[!h] \centering
  \includegraphics[height=4.5cm,width=.45\linewidth]{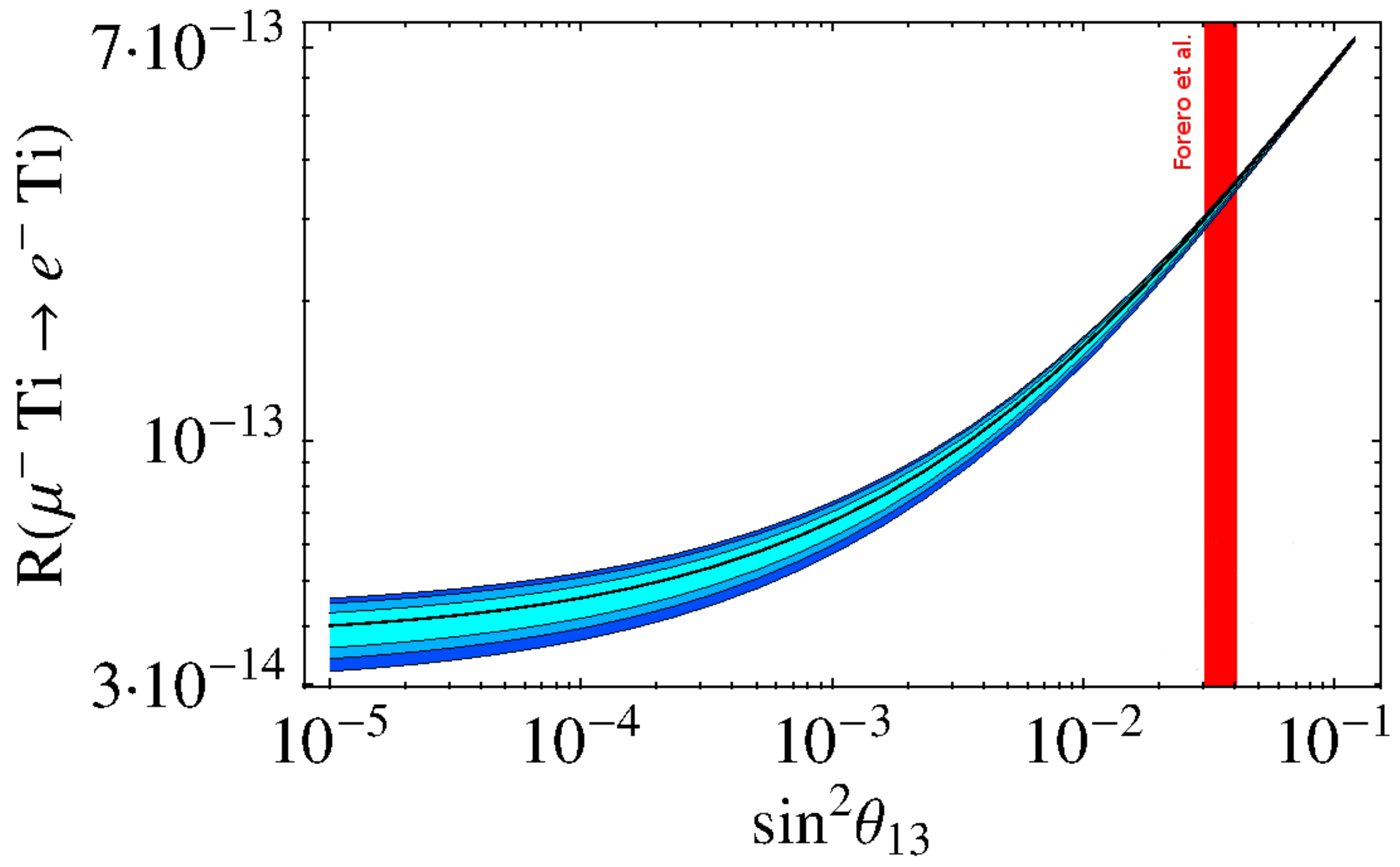}
  \includegraphics[height=4.5cm,width=.45\linewidth]{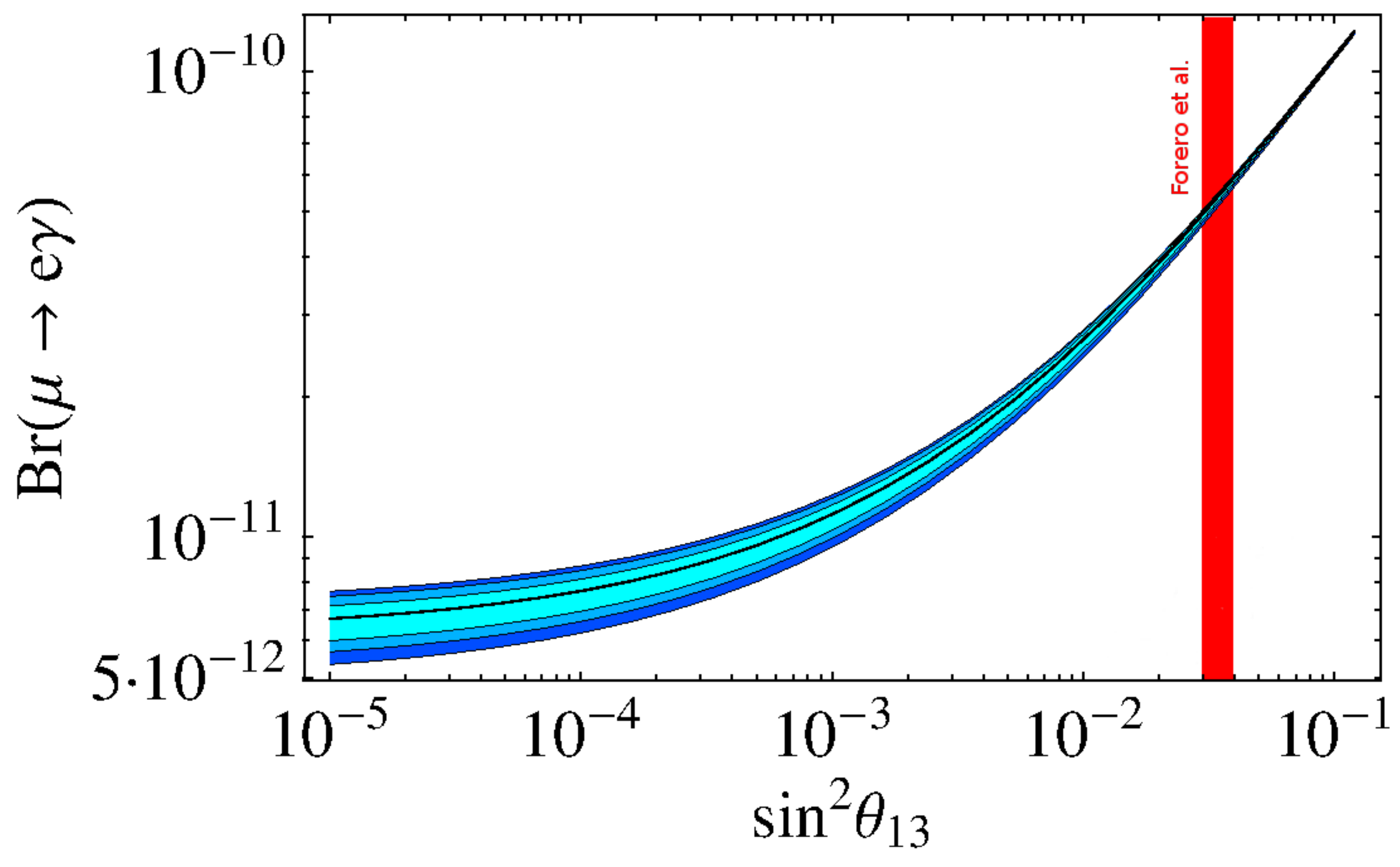}
 \caption{\label{fig:mueg2} LFV branching ratios in the supersymmetric
   inverse seesaw model of neutrino mass (see text).}
\end{figure}

\begin{figure}[!h] \centering
 \includegraphics[height=4.5cm,width=.45\linewidth]{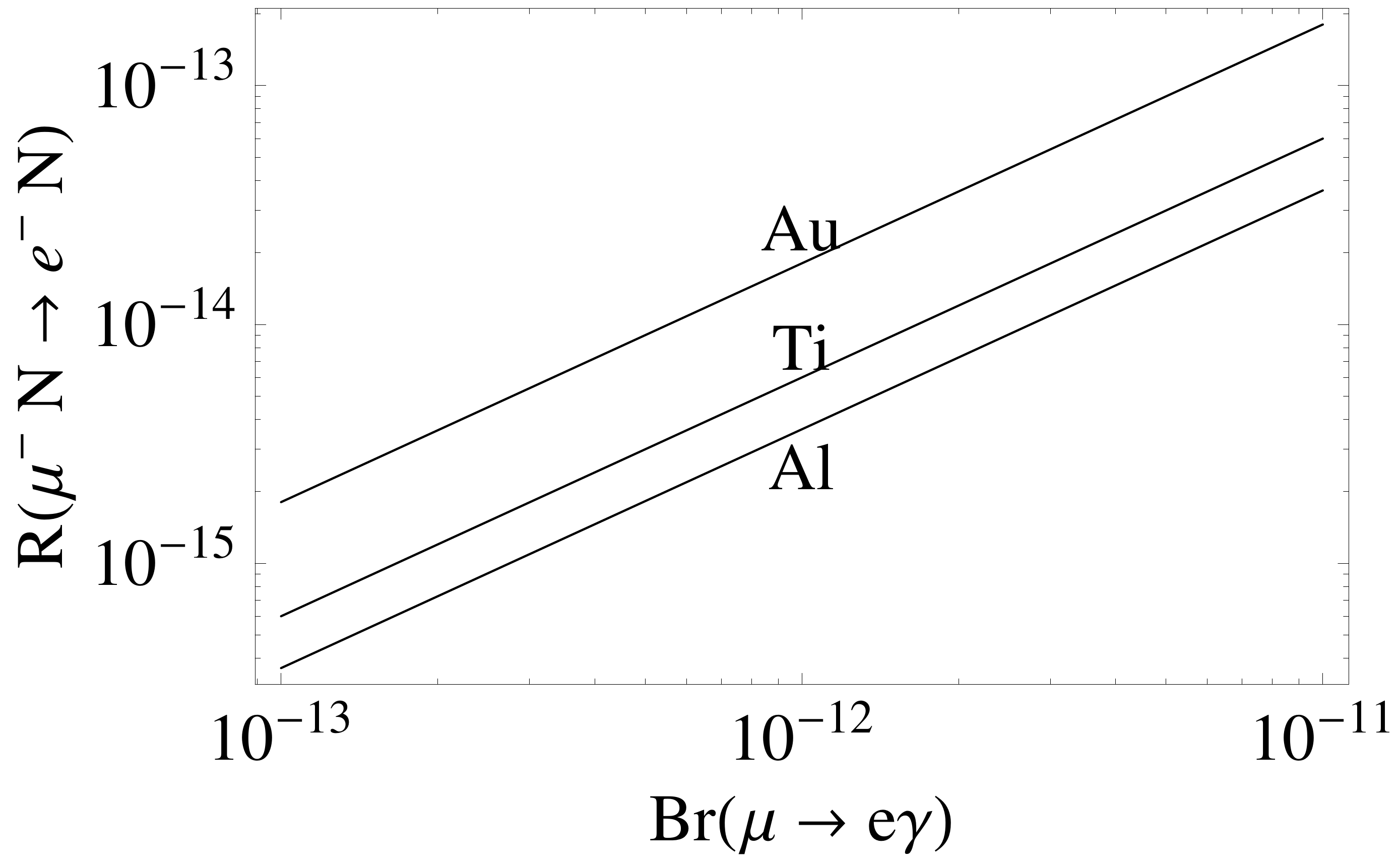}
\caption{\label{fig:mueg3} Correlation between \(Br(\mu\to e\gamma)\)
  and muon-electron conversion in nuclei from Ref.~\cite{Deppisch:2005zm}.}
\end{figure}

Here we consider the rates for the $\mu^-\to e^-\gamma$ decay in the
framework of the supersymmetric inverse seesaw
model~\cite{Deppisch:2004fa,Deppisch:2005zm}.
Fig.~\ref{fig:mueg2} displays the dependence of the branching ratios
for $\mu^--e^-$ conversion in Ti (left) and $\mu^-\to e^-\gamma$
(right) with the small neutrino mixing angle $\theta_{13}$, for
different values of $\theta_{12}$ (black curve: $\theta_{12}$ best fit
value, blue bands denote $2\sigma,~3\sigma,~4\sigma$ confidence
intervals for the solar mixing angle $\theta_{12}$).  The inverse
seesaw parameters are given by: $M=1$~TeV and $\mu=30$~eV.  The light
neutrino parameters used are from \cite{maltoni:2004ei}, except for
$\theta_{13}$ which is varied as shown in the plots.  The vertical
lines indicate the $\sin^2\theta_{13}$ values indicated by recent
experiments, as well as by the global fit in
Ref.~\cite{Tortola:2012te}.  These \lfv rates may be testable in the
new generation of upcoming
experiments~\cite{Kurup:2011zza,Kutschke:2011ux}.  For large \(M\) the
estimates recover those of the standard supersymmetric seesaw.

\subsection{Neutral heavy lepton versus supersymmetric exchange}
\label{sec:neutral-heavy-lepton-1}

The novel feature present in low-scale models and not in the minimal
seesaw is the possibility of enhancing \(Br(\mu\to e\gamma)\) and
other tau decays with \lfv due to neutral heavy lepton (right-handed
neutrino) versus supersymmetric lepton exchange. 
In the case where \(M\) is low, around TeV or so this happens even in
the \emph{absence} of supersymmetry.
In this region of parameters the model also gives rise to large
estimates for the nuclear $\mu^--e^-$ conversion, depicted in
Fig.~\ref{fig:Diagrams}.  The latter fall within the sensitivity of
future experiments~\cite{Kurup:2011zza}. Note that large \lfv rates
are possible even in the massless neutrino limit. The allowed lepton
flavour and CP violation rates are, in fact, unsuppressed by the
smallness of neutrino
masses~\cite{bernabeu:1987gr,branco:1989bn,rius:1990gk,gonzalez-garcia:1992be,Ilakovac:1994kj}.
Finally, for low enough \(M\) the corresponding heavy leptons could be
searched directly at particle accelerators such as LEP already
did~\cite{Dittmar:1990yg,Abreu:1997pa}. Prospects of LHC detection are
less clear, though they are good in the presence of an extended gauge
boson ``portal'', such as right-handed gauge
bosons~\cite{AguilarSaavedra:2012fu,Das:2012ii}.

\section{Collider tests: probing neutrino flavour mixing at the LHC
}
\label{sec:coll-tests:-prob}

\begin{figure}[!h]
\centering
\includegraphics[clip,height=5cm,width=0.42\linewidth]{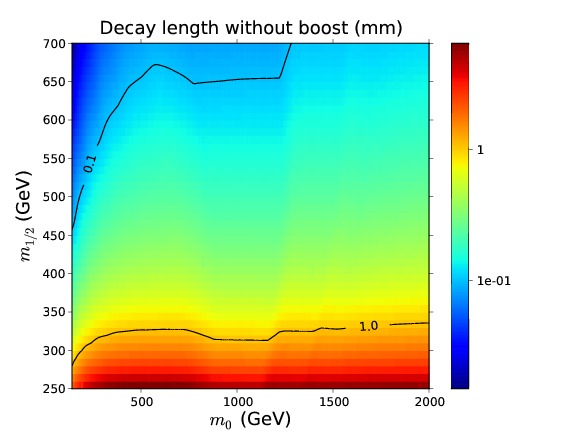}
\includegraphics[clip,height=5cm,width=0.42\linewidth]{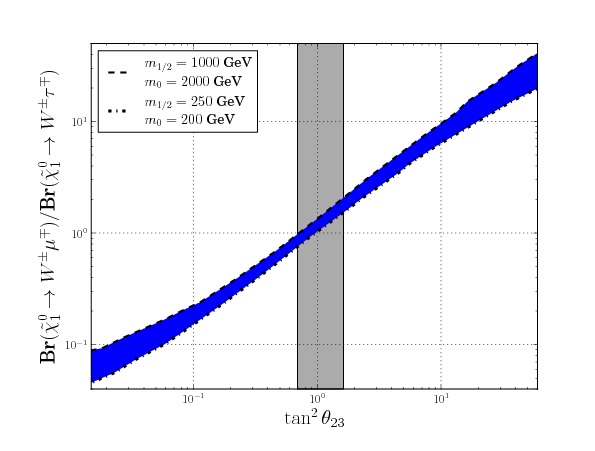}
\caption{LSP decays typically produce displaced vertices inside the
  detector, their semi-leptonic decay branching ratios correlate well
  with the atmospheric mixing
  angle~\cite{DeCampos:2010yu,Porod:2000hv}, whose low energy
  determination is illustrated by the vertical
  band~\cite{Tortola:2012te}.}
     \label{fig:correlation}
\end{figure}

We now turn to the case of low-scale models of neutrino mass. As an
example we consider the case of models where supersymmetry is the
origin of neutrino mass~\cite{Hirsch:2004he,Hirsch:2000ef}, considered
in Sec.~\ref{sec:supersymm-as-orig}.
A general feature of these models is that the lightest supersymmetric
particle (LSP) is unstable, since it is not protected by any
symmetry. In order to reproduce the masses required by current
neutrino oscillation data, the LSP is typically expected to decay
inside the detector, leaving a displaced
vertex~\cite{magro:2003zb,decampos:2007bn,DeCampos:2010yu,deCampos:2012pf}
as seen in the left panel in Fig.~\ref{fig:correlation}.  More
strikingly, its decay properties correlate with neutrino mixing
angles, as seen in the right panel. For example, if the LSP is the
lightest neutralino, it is expected to have the same decay rate into
muons and taus, since the observed atmospheric angle, is relatively
close to
$\pi/4$~\cite{Porod:2000hv,romao:1999up,mukhopadhyaya:1998xj}.
This opens the tantalizing possibility of testing neutrino mixing at
high energy accelerators, like the LHC and the "International Linear
Collider" (ILC) and constitutes a smoking gun signature of this
proposal that for sure will be tested.
This possibility also illustrates the complementarity of accelerator
and non-accelerator approaches in elementary particle physics. Before
closing this discussion we mention a recent attempt to introduce a
flavour symmetry to the bilinear R-parity violation
scheme~\cite{Bazzocchi:2012ve}.

\section{Implications of a ``large'' reactor angle}
\label{sec:non-zero-reactor}

Recently reactor experiments Double CHOOZ~\cite{Abe:2011fz}, Daya
Bay~\cite{An:2012eh} and RENO~\cite{Ahn:2012nd} have published 
\begin{eqnarray}
\sin^2 2\theta_{13}&=& 0.092\pm 0.016({\rm stat})\pm0.005({\rm syst}) \qquad {\rm at}\, 5.2\sigma  \qquad {\rm (Daya Bay)}\label{best131}\\
\sin^2 2\theta_{13}&=& 0.113 \pm 0.013({\rm stat}) \pm 0.019({\rm syst}) \qquad {\rm at}\, 4.9\sigma \qquad {\rm (RENO)}\label{best132}
\end{eqnarray}
with similar results recently presented at the Neutrino 2012
conference in Kyoto.

Here we argue that the TBM \emph{ansatz} can  still be taken as a good first
order approximation.  As we discussed in the introduction, in order to
have the TBM mixing pattern we need to break separately our flavour
group (for instance $A_4$) into $Z_3$ in the charged sector and into
$Z_2$ in the neutrino sector.  Therefore the flavour group is completely
broken. Since the flavour symmetry leading to the TBM \emph{ansatz} is in
general broken we expect deviations from TBM which, in particular,
could generate a nonzero reactor angle.

However many models having TBM at leading order are ruled out because
of the recent results which indicate that $\sin\theta_{13}\sim
\lambda_C$ where $\lambda_C\approx 0.2$. In fact in general we expect
that next--to--leading order terms give corrections of the same order
$\delta_\theta$ to the three angles $\theta_{13}$, $\theta_{12}$ and
$\theta_{23}$. Assuming $\delta_\theta\sim \lambda_C$ we have
\begin{eqnarray}
\sin^2 2\theta_{13}&=& 0.087  \qquad {\rm for} \,\delta_\theta = 0.15\\
\sin^2 2\theta_{13}&=& 0.152  \qquad {\rm for} \,\delta_\theta = 0.20
\end{eqnarray}
close to the best fits in (\ref{best131}) and (\ref{best132}). However
the deviations of the solar mixing from its trimaximal values
$\sin^2\theta_{12}^{TBM}\equiv 1/3$ will be too large if we take
$\delta_\theta  = 0.15 \sim \lambda_C$, namely
\begin{eqnarray}
\sin^2 (\theta_{12}^{TBM}+\delta_\theta)&=& 0.48 \quad(0.38 \, @3\sigma) \\
\sin^2 (\theta_{12}^{TBM}-\delta_\theta)&=& 0.20  \quad(0.27 \, @3\sigma). 
\end{eqnarray}
While this poses no problem for the BMV model, which does not predict
the solar angle~\cite{Babu:2002dz}, it in principle rules out most
TBM schemes.
Indeed, most extensions of TBM models which allow for a large reactor
angle also predict a deviation of the atmospheric, see for instance
\cite{Antusch:2011ic,Ishimori:2012fg}, and/or the, by now
well-measured, solar mixing angle from their TBM values. Therefore one
of the most relevant theoretical and experimental questions is to
evaluate the extent to which solar and atmospheric mixing angles
\emph{deviate} from their TBM values.

Still, not all TBM models proposed in the past are excluded, for
example in the model of Ref.\,\cite{Lin:2009bw}, based on $A_4$, large
reactor angle $\theta_{13}\sim \lambda_C$ has been obtained with
deviation of $\theta_{12}^{TBM}$ of order of $\lambda_C^2$ in
agreement with data.
There are other examples in the literature of models where such
deviations are negligible, despite the relatively large reactor angle
value, see for instance
\cite{Morisi:2011pm,Araki:2011wn,Toorop:2011jn,Bazzocchi:2011ax,King:2012vj}.

Many alternative \emph{ansatze} have been considered in order to
circumvent this problem. An interesting possibility is that the
leading order neutrino mass matrix is not diagonalized by the TBM
\emph{ansatz}, but rather by the \emph{bi-maximal} one (where both
solar and atmospheric mixing angles are maximal from the start)
\cite{Altarelli:2009gn} or simply
\emph{bi-large}~\cite{Boucenna:2012xb}, or by the \emph{golden ratio}
scheme \cite{Ding:2011cm,King:2012vj}.
Clearly a ``large'' reactor angle will not only act as a ``portal'' to
a new world of CP violation in the lepton sector, but may also shed
light into the flavour problem, one of the deepest puzzles to our
current theories of matter.

\begin{acknowledgement}

  This work was supported by the Spanish MINECO under grants
  FPA2011-22975 and MULTIDARK CSD2009-00064 (Consolider-Ingenio 2010
  Programme), by Prometeo/2009/091 (Generalitat Valenciana), by the EU
  ITN UNILHC PITN-GA-2009-237920. S.M. acknowledges Juan de la Cierva
  contract.
\end{acknowledgement}


\end{document}